\documentclass[aps,pre,superscriptaddress]{revtex4}
\usepackage[dvips]{graphics}
\usepackage{graphicx}
\usepackage{amsfonts}
\usepackage{amssymb}
\usepackage{amsmath}
\usepackage{subfigure}

\begin{document}

\title{ Solitary Waves in Discrete Media with Four Wave Mixing }
\author{R.L. Horne}
\affiliation{Department of Mathematics, Florida State University, Tallahassee, FL 32306-4510}
\email{horne@math.fsu.edu}

\author{P.G.\ Kevrekidis }
\affiliation{Department of Mathematics and Statistics, University of Massachusetts,
Amherst MA 01003-4515, USA}
\email{kevrekid,whitaker@math.umass.edu}
\author{N. Whitaker}
\affiliation{Department of Mathematics and Statistics, University of Massachusetts,
Amherst MA 01003-4515, USA}

\begin{abstract}
In this paper, we examine in detail the principal branches of solutions
that arise in vector discrete models with nonlinear inter-component
coupling and four
wave mixing. The relevant four branches of solutions consist of two
single mode branches (transverse electric and transverse magnetic)
and two mixed mode branches, involving both components (linearly 
polarized and elliptically polarized). These solutions are obtained
explicitly and their stability is analyzed completely in the anti-continuum
limit (where the nodes of the lattice are uncoupled), illustrating the
supercritical pitchfork nature of the bifurcations that give rise to
the latter two, respectively, from the former two. Then the branches
are continued for finite coupling constructing a full two-parameter 
numerical bifurcation diagram of their existence. Relevant stability
ranges and instability regimes are highlighted and, whenever unstable,
the solutions are dynamically evolved through direct computations to 
monitor the development of the corresponding instabilities. Direct
connections to the earlier experimental work of Meier {\it et al.}
[Phys. Rev. Lett. {\bf 91}, 143907 (2003)] that motivated the present
work are given. 
\end{abstract}


\maketitle

\section{Introduction}
 
Recently, nonlinear Hamiltonian lattice dynamical systems with a large number of degrees of freedom
have become a focal point for a variety of application areas \cite{reviews}.
Several diverse physical contexts in which such models (typically discrete in space
and continuous in the evolution variable) arise are (i) spatial dynamics of optical beams in coupled waveguide 
arrays arising in nonlinear optics \cite{reviews1}, (ii) temporal evolution of Bose-Einstein condensates (BECs) 
in optical lattices in soft-condensed matter physics \cite{reviews2} and (iii) DNA double strand in biophysics 
\cite{reviews3}. 

A prototypical model, applicable to different degrees of approximation,
in all of the above contexts is the so-called discrete nonlinear 
Schr{\"o}dinger (DNLS) equation. This model was first proposed in
\cite{b3} and implemented for the first time experimentally in Refs. \cite{b7,b8}. A
systematic presentation of the experimental results 
in optical systems is given in a recent
work \cite{JOSA}; for BEC-related experiments, see \cite{reviews2,morsch}. 
In this model, self-localized excitations (discrete
solitons) are possible as a result of the interplay between the Kerr
nonlinearity and discrete linear coupling. Many properties of optical
discrete spatial solitons have been systematically explored in theory
and experiment, including generalizations to diffraction management \cite
{b9,Jena}, diffraction-managed solitons \cite{b10}, and soliton transport 
and gating \cite{Eugenieva,dnc2,hector,bright}.

On the other hand, a topic that has received considerably less
attention has been the study of vector analogs of the DNLS 
equation. A number of studies have addressed the existence
and stability of diverse families of solitary waves/localized states 
in the DNLS equation (see \cite{c7,c13,epjd}). These issues have also been 
studied for both cubic and quadratic nonlinearities \cite{c13,c18} in 
one and two dimensions \cite{c7,c19,hudock}. 
However, the first experimental realization for
such a system occurred rather recently \cite{meier}
(see also the very recent paper \cite{meier2}). This work sheds new light
on the theoretical investigations of this topic, in that the relevant
model for the corresponding AlGaAs waveguide array experiment
included four wave mixing (FWM) terms that were rarely included 
in previous theoretical studies. Subsequent studies (see 
\cite{molina1,molina2}) addressed a number of specific properties of
the model system put forward in \cite{meier}. Such properties related to 
switching, instability-induced amplification, modulational instability 
as well as
an analysis of the energy barrier between stable discrete solitary waves centered on 
a lattice site versus two lattice sites. 
 

The focus of the present study is to extend the analysis of the vector NLS 
lattice system incorporating FWM which is relevant to the experiments of \cite{meier}.
More specifically, we provide a systematic analysis of the 
existence and stability 
requirements of the principal solitary wave modes of the system.   
In so doing, we have found that the concept of the anti-continuum limit 
(in which the coupling between waveguides is absent) provides
a powerful tool for the study of the existence and stability concerning 
these solutions. Subsequently, continuations in the coupling parameter
are used to construct a full two-parameter bifurcation diagram
that gives the complete picture of stability for both one-component as well
as ``mixed'' mode solutions (involving both components). We note that 
all of the results obtained via our dimensionless model can be directly 
placed in the context of 
the experimental results obtained in Refs. \cite{meier,meier2}. 

The remainder of the paper is organized in the following way. In section II, 
we present the theoretical model of interest as well as proposing a setup 
that allows for analysis of the existence and stability of the model's solutions. 
In section III, we fully analyze the relevant solutions and their stability 
in the anti-continuum limit (i.e., zero coupling between adjacent waveguides). 
In section IV, we proceed numerically to determine solutions to the model 
for nonzero coupling and observe how the bifurcation points and stability
features are modified for this case. The instability
regions obtained in section IV are then monitored through direct
numerical experiments in section V. Finally, in section VI, we
summarize our findings and present our conclusions.

\section{Model and Setup}

Following \cite{meier}, we will use the dimensionless model
\begin{eqnarray}
i \dot{a}_n &=&-a_n-\epsilon \left(a_{n+1}+a_{n-1}\right) - 
\left(|a_n|^2 + A |b_n|^2\right)a_{n} - B b_n^2 a_n^{\star}
\label{req1}
\\
 i \dot{b}_n &=& b_n-\epsilon \left(b_{n+1}+b_{n-1}\right) - 
\left(|b_n|^2 + A |a_n|^2\right)b_{n} - B a_n^2 b_n^{\star}.
\label{req2}
\end{eqnarray}
In these equations, $a_n$ and $b_n$ are the appropriately 
normalized, slowly-varying, complex field envelopes for the transverse
electric (TE) and transverse magnetic (TM) polarized waves respectively.
The constants $A$ and $B$ are respectively associated with
the cross-phase modulation (XPM) and four-wave mixing (FWM)
and were evaluated in \cite{meier} to be approximately equal
to $A \simeq 1$ and $B \simeq 1/2$. Wherever possible,
the results obtained in this analysis will be given for general $A$ and $B$,
even though the above values have been used in the specifics of
our numerical computations. The overdot in Eqs. (\ref{req1})-(\ref{req2})
denotes differentiation with respect to the evolution variable (which
is $z$ in this case) and the $\star$ denotes the complex conjugate. 
For the properties of the waveguide array and incident light
used in the experiment (see \cite{meier} for details), 
the dimensionless power 
\begin{eqnarray}
P = \sum_n (|a_n|^2 + |b_n|^2)
\label{req3a}
\end{eqnarray}
is connected with its dimensional analog $P_d$ (measured in Watts)
according to $P_d \simeq 56.4 P$. Furthermore, the 
dimensionless coupling $\epsilon$ used in \cite{meier} was
$\epsilon \simeq 0.921$. This is crucial since it serves to highlight
differences between the results obtained in the computational analysis
of \cite{meier} and the generalization of the corresponding results 
presented herein. More generally, we take $\epsilon$ to be a free parameter
in our analysis. We note that $\epsilon$ is related to the dimensional coupling
constant, $\kappa$, given in \cite{meier}, via $\epsilon = 2.741\kappa$.
Here, $\kappa$ is measured in $\mbox{mm}^{-1}$.

We look for stationary solutions in the form: $a_n=\tilde{a}_n e^{i q z}$
and $b_n=\tilde{b}_n e^{i q z}$ and subsequently dropping the tildes,
we obtain the corresponding stationary equations:
\begin{eqnarray}
(q-1) a_n-\epsilon \left(a_{n+1}+a_{n-1}\right) - 
\left(|a_n|^2 + A |b_n|^2\right)a_{n} - B b_n^2 a_n^{\star} &=& 0
\label{req3}
\\
(q+1) b_n-\epsilon \left(b_{n+1}+b_{n-1}\right) - 
\left(|b_n|^2 + A |a_n|^2\right)b_{n} - B a_n^2 b_n^{\star} &=& 0.
\label{req4}
\end{eqnarray}
In this context, $q$ is an additional free parameter, namely the propagation 
constant ($q$ is also used as a free parameter for the computations presented
in \cite{meier}). Our parameter $q$ is related to the corresponding dimensional
propagation constant, $\delta$, given in \cite{meier}, by $\delta = 0.3648q$. 
Here, $\delta$ is measured in $\mbox{mm}^{-1}$.

It is worth pointing out that for Fig. 1 in Ref. \cite{meier}, the range
of parameters used in their computations appear to correspond to the case of
$q > 3$. This is verified by our analysis below. Again, this is highlighted 
to account for the disparities of some of the conclusions obtained in 
\cite{meier} with the ones their generalizations 
presented in the following sections.

To perform linear stability computations, provided that an
exact solution $(a_n^0,b_n^0)$ 
of the stationary Eqs. (\ref{req3})-(\ref{req4})
has been obtained (either analytically in the anti-continuum
limit, or numerically for $\epsilon \neq 0$), we use the perturbation
ansatz:
\begin{eqnarray}
a_n &=& a_n^0 + \delta \left(c_n e^{-i \omega z} + d_n e^{i \omega^{\star} z}
\right)
\label{req5}
\\
b_n &=& b_n^0 + \delta  \left(f_n e^{-i \omega z} + g_n e^{i \omega^{\star} z}
\right).
\label{req6}
\end{eqnarray}
In this ansatz, $\delta$ is a (small) formal expansion parameter and
$\omega$ is the eigenfrequency of the 
perturbation eigenvector
$(c_n,d_n^{\star},f_n,g_n^{\star})^T$ (where the superscript denotes
the transpose). Then, upon lengthy but straightforward calculation, 
one obtains the linearized matrix eigenvalue problem to O$(\delta)$. 
These dynamical equations have the form:
\[
\omega \left( 
\begin{array}{c}
c_{n} \\ 
d_{n}^{\star} \\ 
f_{n} \\ 
g_{n}^{\star}
\end{array}
\right) ={\bf L}\cdot \left( 
\begin{array}{c}
c_{n} \\ 
d_{n}^{\star} \\ 
f_{n} \\ 
g_{n}^{\star}
\end{array}
\right) ,\newline
\]
where 
\[
{\bf L}=\left( 
\begin{array}{cccc}
L_{11} & L_{12} & L_{13} & L_{14} \\ 
L_{21} & L_{22} & L_{23} & L_{24} \\ 
L_{31} & L_{32} & L_{33} & L_{34} \\ 
L_{41} & L_{42} & L_{43} & L_{44}
\end{array}
\right) .\newline
\]
The $N \times N$ (where $N$ is the size of the lattice) blocks
of the linearization matrix are given by:
\begin{eqnarray}
L_{11} &=& (q-1)-\epsilon (\Delta_2+2) - 2 |a_n^0|^2 - A |b_n^0|^2
\label{req7}
\\
L_{12} &=& -(a_n^0)^2 - B (b_n^0)^2
\label{req8}
\\
L_{13} &=& -A a_n^0 (b_n^0)^{\star}  - 2 B (a_n^0)^{\star} b_n^0
\label{req9}
\\
L_{14} &=& -A a_n^0 b_n^0 
\label{req10}
\\
L_{21} &=& - L_{12}^{\star}
\label{req11}
\\
L_{22} &=& -L_{11}
\label{req12}
\\
L_{23} &=& - L_{14}^{\star}
\label{req13}
\\
L_{24} &=& - L_{13}^{\star}
\label{req14}
\\
L_{31} &=& L_{13}^{\star}
\label{req15}
\\
L_{32} &=& L_{14}
\label{req16}
\\
L_{33} &=&  (q+1)-\epsilon (\Delta_2+2) - 2 |b_n^0|^2 - A |a_n^0|^2
\label{req17}
\\
L_{34} &=& - (b_n^0)^2 - B (a_n^0)^2
\label{req18}
\\
L_{41} &=& -L_{14}^{\star}
\label{req19}
\\
L_{42} &=& - L_{13}
\label{req20}
\\
L_{43} &=& - L_{34}^{\star}
\label{req21}
\\
L_{44} &=& - L_{33}
\label{req22}
\end{eqnarray}
In the above, the short-hand notation $(\Delta_2+2) z_n = z_{n+1} + z_{n-1}$,
invoking the concept of the discrete Laplacian $\Delta_2$. With this linearized
matrix eigenvalue problem, we can establish a
criterion for the existence and stability
of solutions to this system by examination of our 
system at the anti-continuum limit
(i.e., when $\epsilon = 0$).

\section{The Anti-Continuum Limit}

\subsection{Existence}

We begin by examining eqs. $(\ref{req3})-(\ref{req4})$ for the case $\epsilon$ = 0. 
Setting $a_n=r_n e^{i \theta_n}$ and $b_n=s_n e^{i \phi_n}$ in this case, one finds:
\begin{eqnarray}
(q-1)-(r_n^2+A s_n^2) - B s_n^2 e^{2 i (\phi_n-\theta_n)} &=& 0
\label{req23}
\\
(q+1)-(s_n^2+A r_n^2) - B r_n^2 e^{-2 i (\phi_n-\theta_n)} &=& 0.
\label{req24}
\end{eqnarray}
From the above, we infer that for solutions to exist in this limit
it is necessary for $2 (\theta_n-\phi_n)$ to be an integer multiple of 
$\pi$, hence, we obtain that
\begin{eqnarray}
\theta_n-\phi_n= k \frac{\pi}{2},
\label{req25}
\end{eqnarray}
with $k \in {\cal Z}$. 

The simplest possible solutions are the ones
that involve only one of the two branches and were hence termed
TE and TM modes respectively in \cite{meier}. The TE solution 
of Eqs. (\ref{req23})-(\ref{req24}) has the form (in the present limit)
\begin{eqnarray}
r_n= \pm \sqrt{q-1} \quad s_n =0
\label{req26}
\end{eqnarray}
and exists only for $q>1$.
On the other hand, the TM mode features
\begin{eqnarray}
r_n=0 \quad s_n= \pm \sqrt{q+1},
\label{req27}
\end{eqnarray}
and is only present for $q>-1$. 

Also, there are two possible mixed mode solutions permitted by the quantization
condition of eq. (\ref{req25}), allowing the exponential of Eqs. 
(\ref{req23})-(\ref{req24}) to be $+1$ or $-1$ respectively. The first
one $(e^{2i(\theta_{n} - \phi_{n})} = 1)$ was characterized as a linearly polarized 
(LP) branch in \cite{meier},
involving in-phase contributions from both the TE and TM components.
In this case, the linear system of Eqs. (\ref{req23})-(\ref{req24})
has the general solution:
\begin{eqnarray}
r_n &=& \pm \sqrt{\frac{(A+B)(q+1)-(q-1)}{(A+B)^2-1}}
\label{req28}
\\
s_n &=& \pm \sqrt{\frac{(A+B)(q-1)-(q+1)}{(A+B)^2-1}}.
\label{req29}
\end{eqnarray}
If $(A+B)^2>1$ (which is the case in the experimentally relevant
setting), this branch only exists for $(A+B) (q+1)>(q-1)$
and $(A+B) (q-1)>(q+1)$ [the sign of the two above
inequalities needs to be reversed for existence conditions
in the case of $(A+B)^2<1$]. Among the two conditions, in the
present setting, the second one is the most ``stringent''
for the case $A=2B=1$, which yields the constraint $q \geq 5$
(while the first condition requires for the same parameters $q \geq -5$).

The second possibility for a mixed mode (stemming from
$e^{2 i (\theta_n-\phi_n)}=-1$ in Eqs. (\ref{req23})-(\ref{req24}))
yields the, so-called, elliptically polarized mode (EP) \cite{meier},
involving a $\pi/2$ phase shift between the TE and TM components.
The relevant amplitudes in this case are:
\begin{eqnarray}
r_n &=& \pm \sqrt{\frac{(A-B)(q+1)-(q-1)}{(A-B)^2-1}}
\label{req30}
\\
s_n &=& \pm \sqrt{\frac{(A-B)(q-1)-(q+1)}{(A-B)^2-1}}.
\label{req31}
\end{eqnarray}
The condition for this branch to exist if $(A-B)^2<1$
is $q-1 \geq (A-B) (q+1)$ and $q+1 \geq (A-B) (q-1)$
(once again the signs should be reversed if $(A-B)^2>1$).
In this case, the first condition is more constraining
than the second, imposing for our
special case of interest herein ($A=2B=1$) $q \geq 3$
(while the second condition only requires $q \geq -3$).

\subsection{Stability} 

We can now turn to the study of the stability of the corresponding
branches via the anti-continuum limit ($\epsilon = 0$). The major advantage of
this limit is that all the blocks of the linearization
matrix $L$ now become diagonal. Furthermore, the structures
that we consider, per the branches presented above, involve the excitation 
of a {\it single site} (all other sites are inert) with amplitudes 
corresponding to the appropriate 
TE, TM, LP or EP modes. The interesting byproduct of this formulation is
that all inert sites end up yielding purely diagonal entries
in the stability matrix equal to $(q-1)$ in the first (2N-1)$\times$(2N-1)
blocks and $(q+1)$ in the second set of (2N-1)$\times$(2N-1) blocks.
Hence the matrix will have $2N-1$ eigenfrequencies equal to $(q-1)$
and equally many eigenfrequencies of $(q+1)$. The remaining $4 \times 4$
matrix will correspond to the {\it single excited site} of the
lattice and its eigenvalues will yield the eigenfrequencies
pertaining to the corresponding branch of solutions. Hence, by solving
this considerably simpler eigenvalue problem, we can infer the full
stability picture for $\epsilon \neq 0$ from the anti-continuum limit. 
This, we believe, is a major advantage in using this approach.

We note that due to the fact that the power is conserved for our system (see previous
section), two of the relevant eigenvalues of this $4 \times 4$ problem should be identically 0. 
This is confirmed in the computations presented here.

Now, taking the case for the ``TE'' mode in eqs.$(\ref{req7})-(\ref{req22})$, we determine the 
remaining pair of eigenvalues in this case: 
\begin{eqnarray}
\omega_{TE}=\pm \sqrt{(q+1)^2+(A^2-B^2) (q-1)^2-2 A (q^2-1)}.
\label{req32} 
\end{eqnarray}
Examining our model for the experimental case, $A$ = 1, $B$ = 1/2 and $q > 0$, 
we find that this eigenfrequency is real for $q < 5$, while it is imaginary for $q>5$ 
(hence, implying the presence of an instability,
since the corresponding eigenvalue $\lambda=i \omega$ becomes real
and exponential growth occurs $\propto e^{\lambda z}$). Indeed, we can show that
for the TE mode, stability is ensured for $1 < q < 5$. 

For the TM mode, the corresponding expression can be obtained as:
\begin{eqnarray}
\omega_{TM}=\pm \sqrt{(q-1)^2+(A^2-B^2) (q+1)^2-2 A (q^2-1)},
\label{req33}
\end{eqnarray}
and for the case of interest, stability is ensured for $-1 < q < 3$, while
instability ensues (due to a real eigenvalue) for $q > 3$.

For the LP mode, the expression for the corresponding eigenfrequency
is too cumbersome to provide in analytic form (even though such a form
has been obtained) for general $A$ and $B$. Hence, we will restrict 
ourselves to the case of $A=2B=1$, where the relevant eigenfrequency
reads:
\begin{eqnarray}
\omega_{LP}= \pm \frac{2 \sqrt{2}}{5} \sqrt{q^2-25},
\label{req34}
\end{eqnarray}
implying stability for $q \geq 5$ (which coincides with the branch's 
existence region).

Finally, for the EP mode, the expression similarly reads:
\begin{eqnarray}
\omega_{EP} \pm \frac{2 \sqrt{2}}{3} \sqrt{q^2-9},
\label{req35}
\end{eqnarray}
implying, in principle, stability for $q \geq 3$ (again, coinciding with the
solutions domain of existence); see, however, the discussion below for
this branch.

From the above observations, one can put together the complete
bifurcation picture of the four branches and their corresponding
stability in the anti-continuum limit (again, we present this here
for concreteness for $A=2B=1$). More specifically, 
the TE branch exists for $q \geq 1$ and is stable for 
$1 \leq q \leq 5$. For $q>5$, the branch is destabilized as
a new branch emerges, namely the LP branch, through a
pitchfork bifurcation; notice that the TM
component of this branch, per Eq. (\ref{req29}) is exactly zero
at $q=5$, hence it directly bifurcates from the TE branch.
The two branches of this super-critical pitchfork correspond
to the two signs of $s_n$ in Eq. (\ref{req29}). The bifurcating
branch ``inherits'' the stability of the TE branch for 
all larger values of $q$, while
the latter branch remains unstable thereafter.

In a similar development, the TM branch exists for $q \geq -1$
and is stable in the interval $-1 \leq q \leq 3$. However, at 
$q=3$, a new branch (in fact, a pair of branches) 
bifurcates acquiring non-zero $r_n$, beyond
the bifurcation point, as per Eq. (\ref{req30}). This is accompanied
by the destabilization of the TM branch (due to a real eigenvalue)
and the {\it apparent} 
stability of the ensuing EP branch for all values of $q>3$.

\begin{figure}[tb]
\medskip
\centerline{
\includegraphics[width=8cm,angle=0,clip]{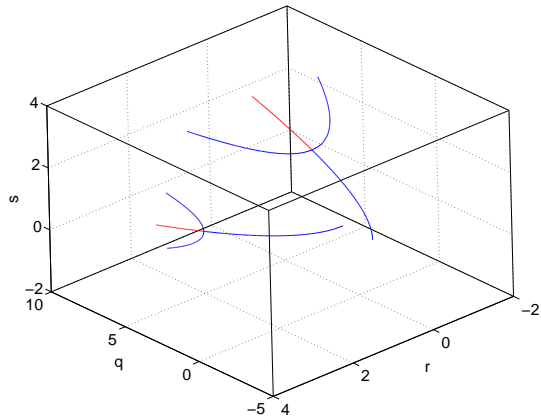}
}
\caption{The full bifurcation diagram for $\epsilon=0$, illustrating the two 
supercritical 
pitchfork bifurcations at $q=3$ and $q=5$. These give rise to the formation
of the EP branch (the former) and the LP branch (the latter). The blue lines
illustrate stable branches, while the red lines show the unstable ones.} 
\label{rfig1}
\end{figure}

Fig. \ref{rfig1} shows how the two supercritical pitchfork bifurcations
are generated as a function of the coordinates $(r_{n}, s_{n}, q)$. We 
can now turn on the coupling and see how the existence and stability of
each branch is affected by varying $q$ and $\epsilon$.

\section{Finite Coupling}

One of the key features that will help us in our analysis of the
finite coupling case consists of determining where the bands of the continuous
spectrum of the linearized problem occurs. This is obtained analytically using 
the plane wave ansatz $a_{n} \sim e^{i (kn - \omega z)}$ and $b_{n} \sim
e^{i (kn - \omega z)}$. Plugging our ansatz into the coupled linearized problem
(essentially, using eqs. $(\ref{req1})-(\ref{req2})$ and 
ignoring the nonlinear terms), the
resulting dispersion relations are  
\begin{eqnarray}
\omega &=& (q-1) - 2 \epsilon \cos(k)
\label{req36}
\\
\omega &=& (q+1) - 2 \epsilon \cos(k).
\label{req37}
\end{eqnarray}
 Hence the continuous spectrum (in the infinite lattice case)
will consist of the frequency intervals $[q-1-2\epsilon,q-1+2\epsilon]$
and $[q+1-2\epsilon,q+1+2\epsilon]$. As may be expected, for 
$\epsilon=0$, these intervals degenerate to two single points 
($\omega=q-1$ and $\omega=q+1$), in consonance with the discussion
of the previous section. But perhaps more importantly, for 
$q-1 < 2 \epsilon$ (equivalently for $\epsilon > (q-1)/2$), the 
continuous spectrum branch will be crossing the origin, leading
to the collision of the eigenvalues with their mirror symmetric
opposites, that will in turn lead to instabilities. Hence, we need
{\it only} consider couplings in the interval $\epsilon \in [0,(q-1)/2]$.

We now numerically examine each of the main four branches, using
one-parameter continuation in the above-mentioned interval of
$\epsilon$'s for different values of $q$ and thus constructing
a two-parameter bifurcation diagram for each branch. Notice that 
the solutions are numerically obtained for each pair of 
$(q,\epsilon)$ by means of a fixed point iteration. Once
convergent to within a pre-set tolerance ($10^{-8}$ typically for
the computations presented herein), the procedure is followed by numerical 
linear stability analysis, obtaining the eigenfrequencies for the linearization
matrix $L$.

\subsection{TE branch}

\begin{figure}[tbp]
\centerline{
\includegraphics[width=8.cm,height=5cm,angle=0,clip]{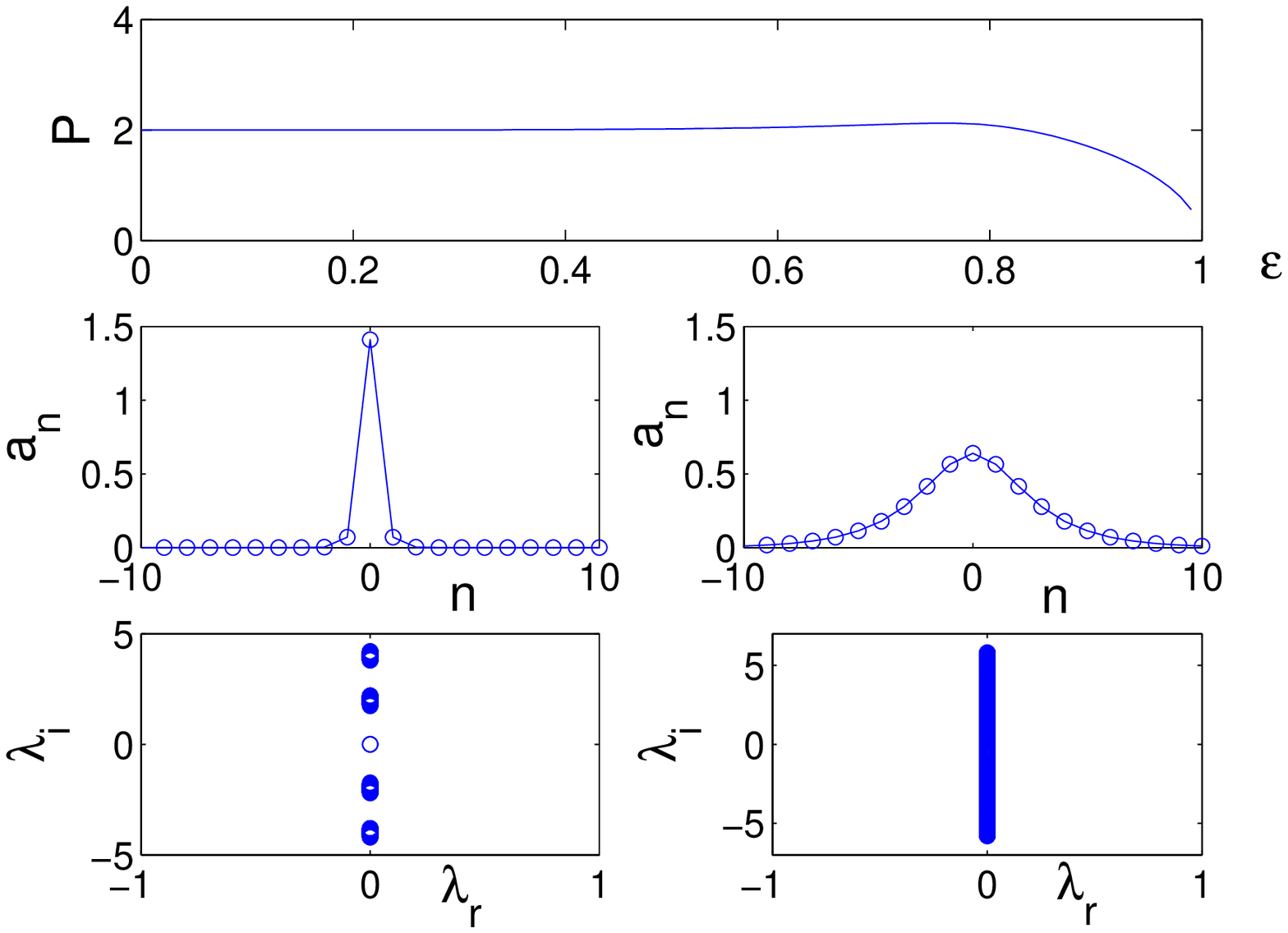}
\includegraphics[width=8.cm,height=5cm,angle=0,clip]{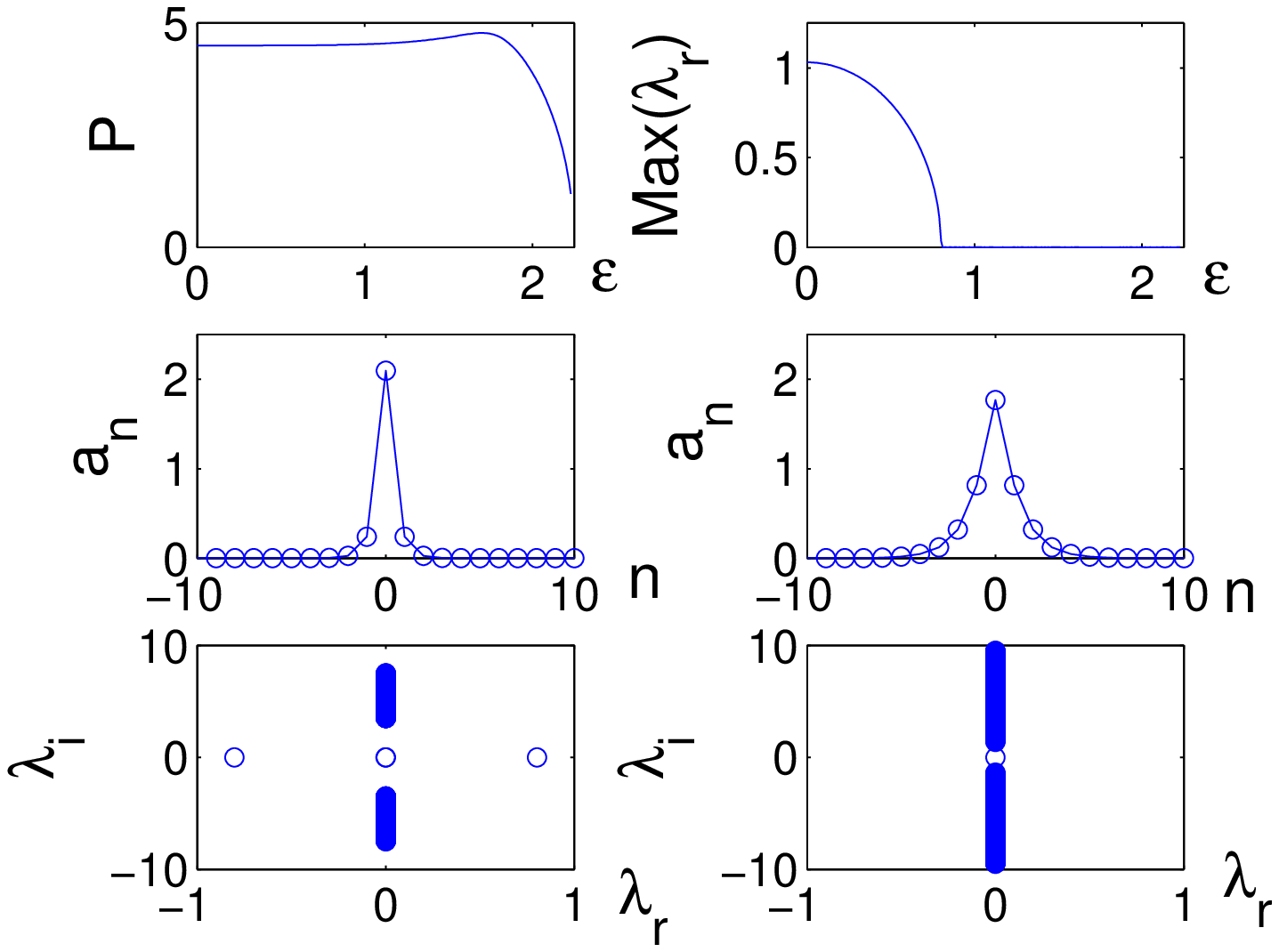}
}
\medskip
\centerline{
\includegraphics[width=8cm,angle=0,clip]{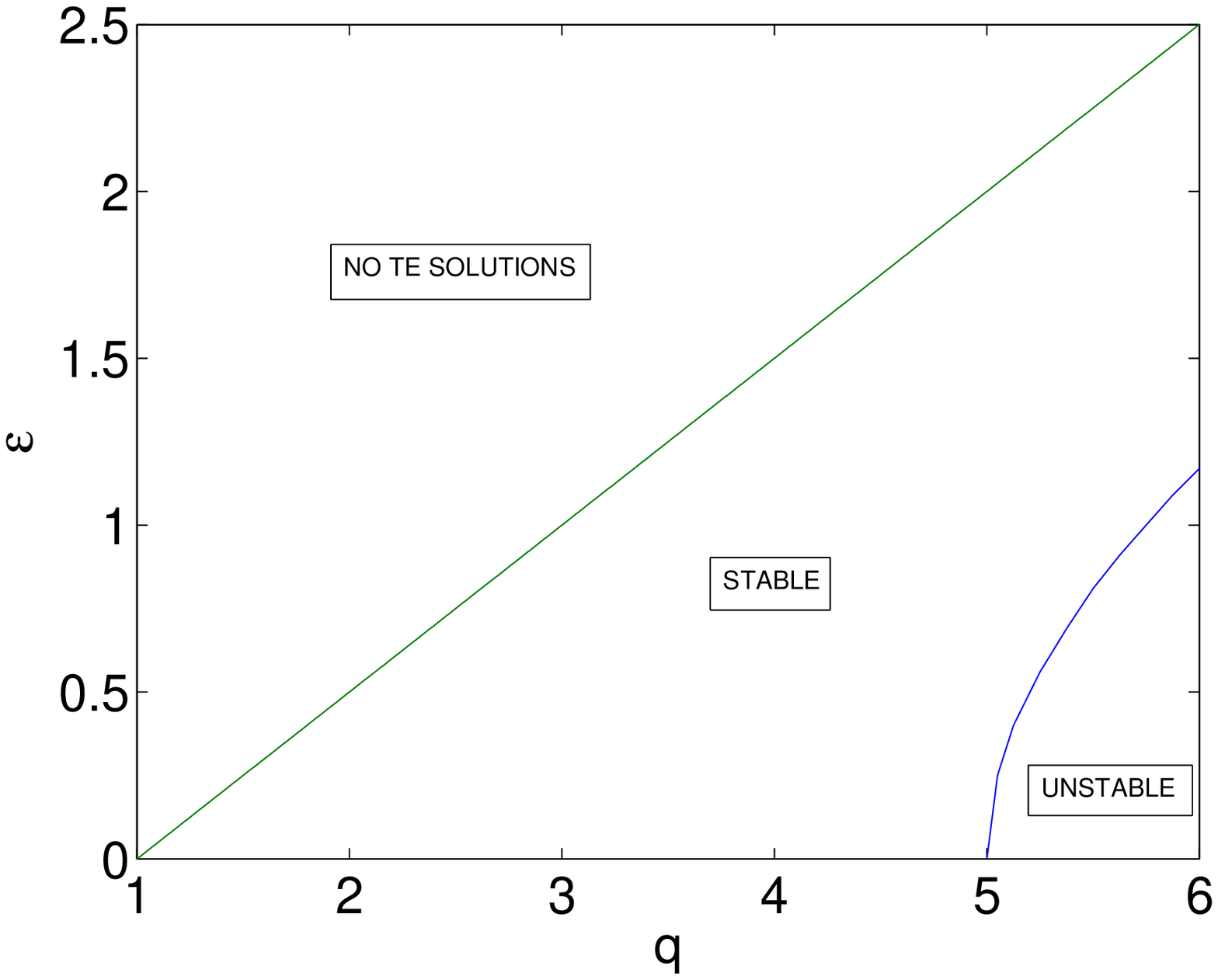}
}
\caption{TE branch: The top left set of panels shows the
details of the TE branch as a function of $\epsilon$ for
$q=3$. The topmost panel shows the norm of the branch as 
a function of $\epsilon$, while the left and right panels
show the solution profile and its spectral plane for
$\epsilon=0.1$ and $\epsilon=0.9$ respectively. The top
right set of panels shows similar details, but for $q=5.5$.
The top right panel in the right case, which is unstable
for small $\epsilon$, shows the most unstable eigenvalue
real part. The middle and bottom panels show the (unstable) solution
and its stability for $\epsilon=0.5$ and the (stable) one for 
$\epsilon=1.5$ respectively.  The bottom panel shows the two
parameter bifurcation diagram of the coupling $\epsilon$ as a function
of $q$ (for details see text). All the relevant existence and stability
regimes have been accordingly labeled. The top left and top right
set of panels can be considered as two ``vertical cuts'' across
this bifurcation diagram for $q=3$ and $5.5$ respectively.} 
\label{rfig2}
\end{figure}

The continuation of the TE branch is detailed in Fig. \ref{rfig2}.
It turns out that especially in the case of this branch, solutions
cannot be obtained for $\epsilon > (q-1)/2$, i.e., the branch
terminates at that point with its amplitude going to zero at
this critical point. Within its region of existence, the branch
has a domain of stability and one of instability. The point of separation
between the two in the anti-continuum limit, studied
previously, was the critical point of $q=5$. For $\epsilon \neq 0$,
the separatrix curve is shown in Fig. \ref{rfig2} and can be well
approximated by the curve $\epsilon_{TE}^c \approx (4 \sqrt{2}/5) \sqrt{q-5}$.
Hence for $q\leq 5$, the solution is stable for all values of
$\epsilon$ in its range of existence ($0<\epsilon<(q-1)/2$), while
for $q \geq 5$, the solution is only stable for $\epsilon_{TE}^c < \epsilon<
(q-1)/2$ and unstable (due to a real eigenvalue pair) for $0< \epsilon <
\epsilon_{TE}^c$.

\subsection{TM branch}

\begin{figure}[tbp]
\centerline{
\includegraphics[width=8.cm,height=5cm,angle=0,clip]{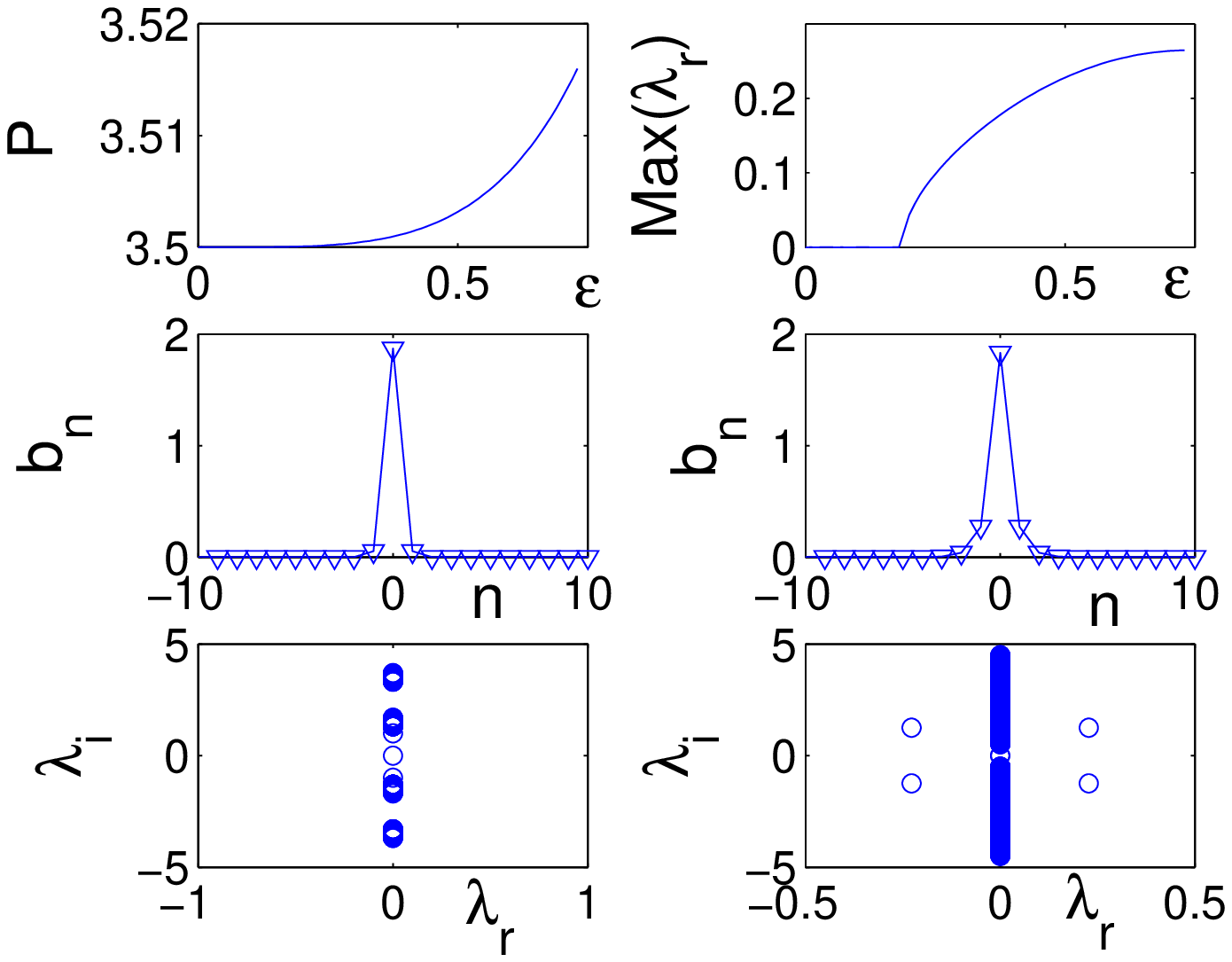}
\includegraphics[width=8.cm,height=5cm,angle=0,clip]{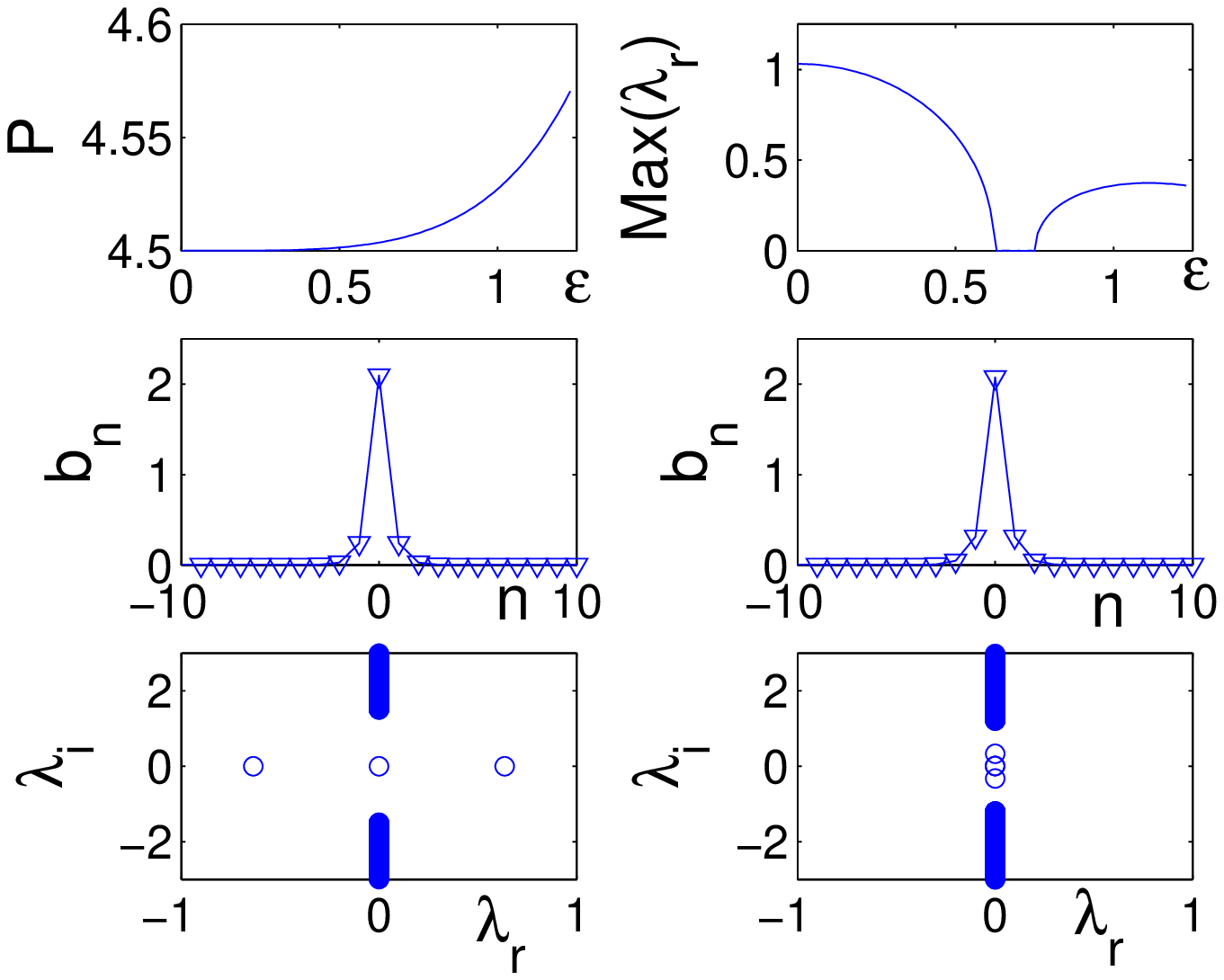}
}
\medskip
\centerline{
\includegraphics[width=8cm,angle=0,clip]{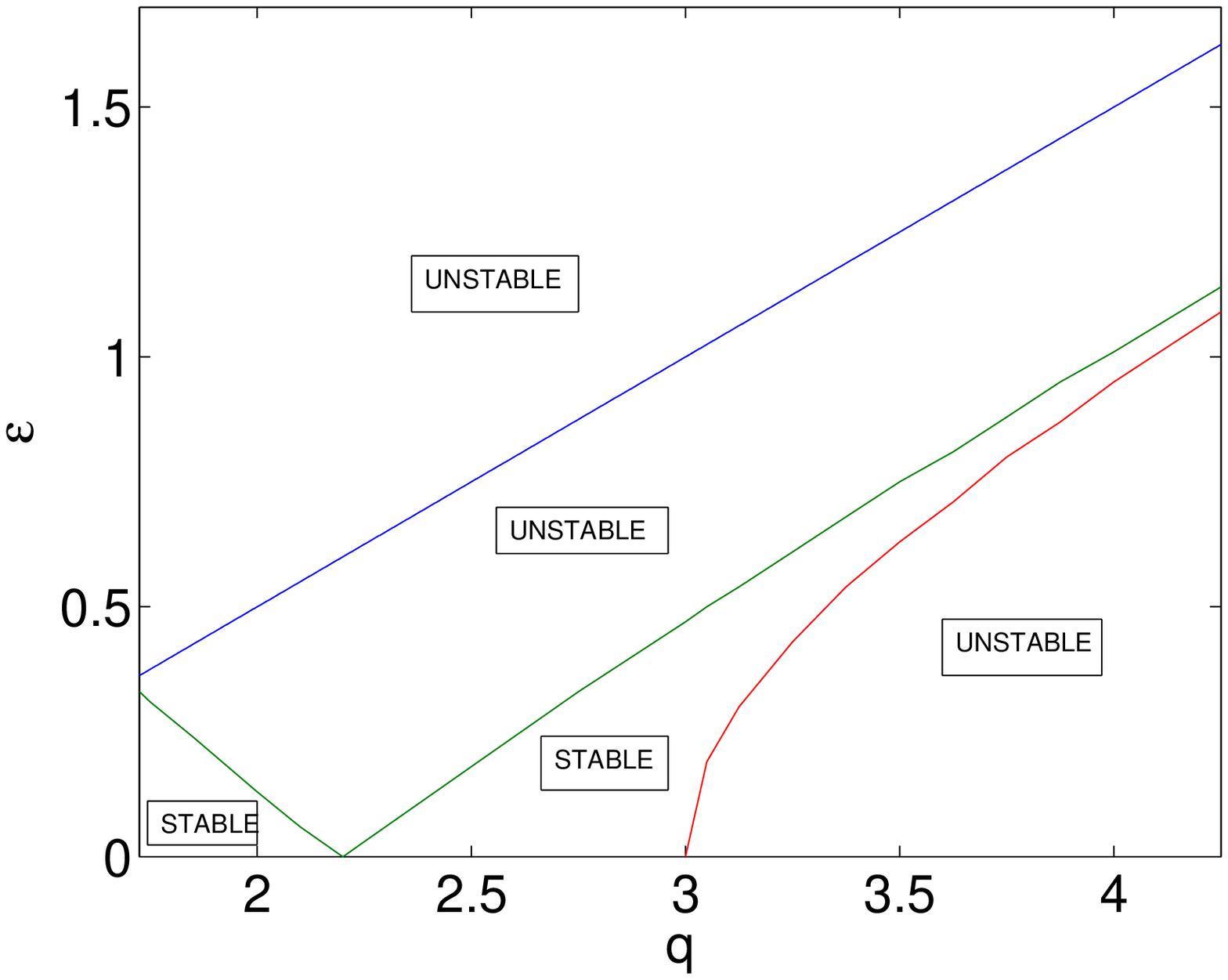}
}
\caption{TM branch: the figures are similar as in Fig. \ref{rfig3},
but now for the TM branch. The top left set of panels shows the
continuation in $\epsilon$ for $q=2.5$, along with the special
cases of $\epsilon=0.1$ (stable) and $\epsilon=0.5$ 
(unstable due to quartet of eigenvalues) in the left
and right middle and bottom panels. The top right set of panels
shows the $q=3.5$ (unstable) case with the middle/bottom panels
corresponding to $\epsilon=0.5$ (unstable due to real
eigenvalue pair)/$\epsilon=0.65$ (stable) respectively.} \label{rfig3}
\end{figure}

The TM branch is considerably more complicated, structurally, than
its TE counterpart. Firstly, it does not disappear beyond the critical
$\epsilon = (q-1)/2$, however, it does become unstable as predicted
previously, hence we will, once again, restrict ourselves to this
parameter range. Furthermore, similarly to the TE branch case, there is an
$\epsilon_{TM}^c$ below which the branch is {\it always} unstable,
whereas for $\epsilon>\epsilon_{TM}^c$, the branch {\it may}
be stable. At the anti-continuum limit, the critical point
for the instability is $q=3$, as discussed previously; for
$q>3$, the critical point is obtained numerically in Fig. \ref{rfig3}.
It can be well approximated (close to $q=3$) by 
$\epsilon_{TM}^c \approx (9/10) \sqrt{q-3}$. 

However, within the range of {\it potential} stability 
($0\leq \epsilon \leq (q-1)/2$ for $q \leq 3$, and 
$\epsilon_{TM}^c \leq \epsilon \leq (q-1)/2$ for 
$q \geq 3$), we observe an additional large region of
instability in the two parameter bifurcation diagram of
Fig. \ref{rfig3}, due to a complex quartet of eigenvalues.
This instability appears to ``emanate'' from the point
with $(q,\epsilon)=(2.2,0)$ in the anti-continuum limit, and
to linearly expand its range as $\epsilon$ increases. Hence,
we will try to understand it at the level of $\epsilon=0$.
What happens at the value of $q=2.2$ in the $\epsilon=0$ limit
is that the point spectrum eigenfrequency of 
Eq. (\ref{req33}) ``collides'' with the continuous spectrum
point of concentration, corresponding to $\omega=q-1$.
However, the eigenvector of this eigenvalue
\footnote{The eigenvector can be found explicitly for the eigenvalues of 
Eq. (\ref{req33}) and in fact is $(-(4 \pm \sqrt{16-(q+1)^2})/(q+1),1,0,0)^T$}
has a Krein signature (see e.g. \cite{aubry,joh,kevkap} for relevant
definitions of this signature) opposite to that of the continuous
band at $\omega=q-1$. Hence, according to Krein theory \cite{mackay}, the
resulting collision leads to the formation of a quartet of eigenvalues
emerging in the complex plane and, in turn, implying the instability of
the TM configuration. As $\epsilon$ is varied from $0$, the continuous
spectrum band grows linearly in $\epsilon$, hence the corresponding 
interval of $q$'s, where this instability is present 
(due to the collision with the opposite Krein sign eigenvalue) 
also grows at the
same rate. Along the same vein, it is worth pointing out that the line
of this instability threshold and that of $\epsilon=(q-1)/2$ are parallel.

\subsection{LP branch}

\begin{figure}[tbp]
\centerline{
\includegraphics[width=8cm,angle=0,clip]{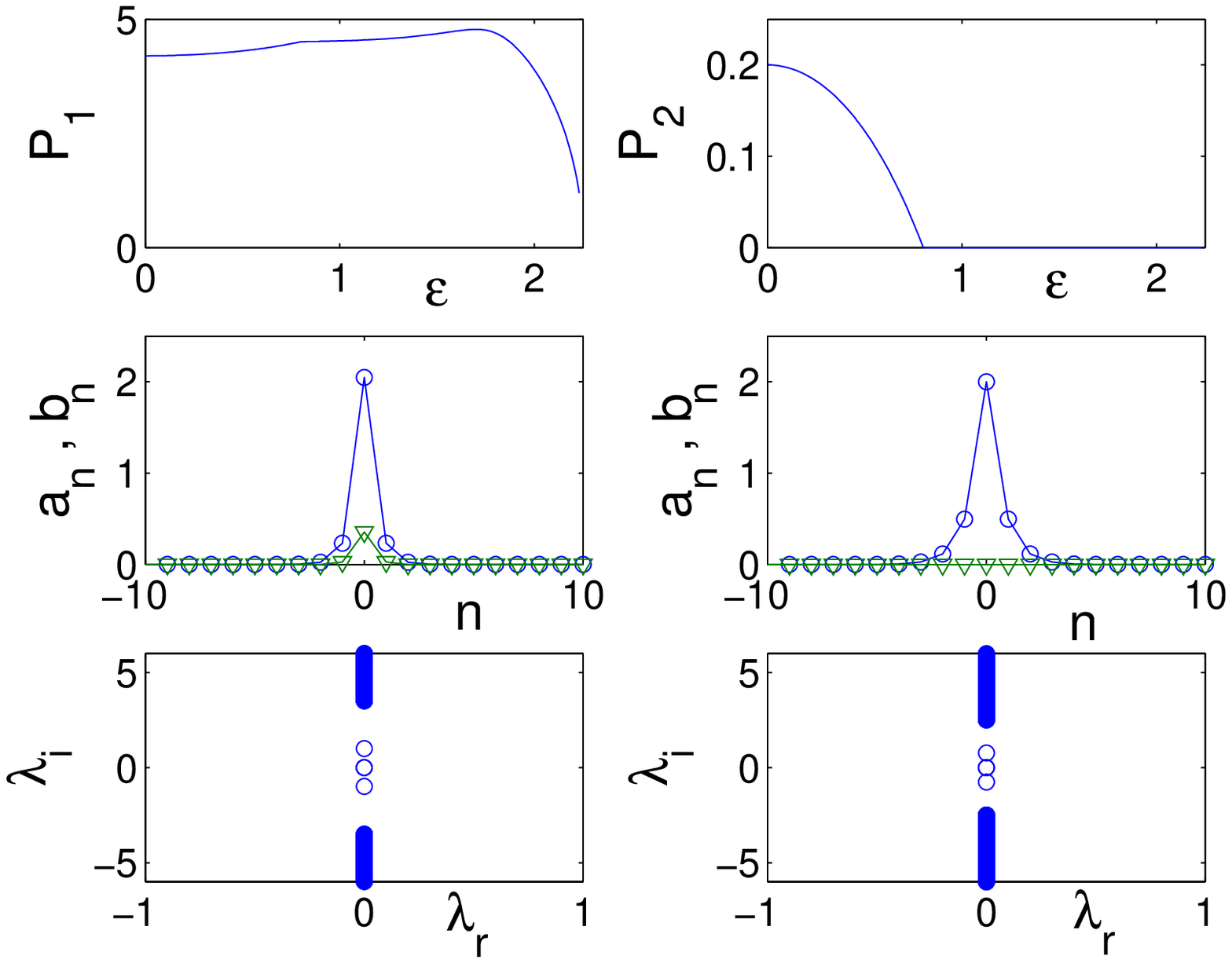}
}
\caption{LP branch: The top left and top right panel shows the 
dependence on $\epsilon$ of the respective powers
of the two components $P_1$ and $P_2$ ($P=P_1+P_2$), for 
$q=5.5$. The middle and bottom panel
panel show the profile and the linear stability of the
LP mode while it is still present ($\epsilon=0.5$; left panels)
and of the TE mode that it degenerates to, beyond $\epsilon_{TE}^c$ 
($\epsilon=1$; right panels).} \label{rfig4}
\end{figure}

The mixed LP branch, involving in-phase contributions of the
TE and TM has been continued for various values of $q>5$,
as a function of $\epsilon$. Recall that this 
branch emerges through a supercritical pitchfork
as $q$ is varied for for fixed $\epsilon$ (see Fig. \ref{rfig2}).
In an exactly analogous way, if $\epsilon$ is varied 
for a fixed $q>5$, the branch appears to terminate 
at $\epsilon=\epsilon_{TE}^c$ through a subcritical pitchfork.
Hence, the LP branch only arises for $0<\epsilon<\epsilon_{TE}^c$
and $q>5$ and it is stable throughout its interval of existence.
For $\epsilon > \epsilon_{TE}^c$, this branch ``degenerates''
back into the stable (for such $\epsilon$'s) TE mode.
This is clearly illustrated in Fig. \ref{rfig4} that shows
the continuation of the LP branch as a function of $\epsilon$
for a fixed $q=5.5$. The top panels show the norms of the
two components illustrating that the second component is
absent for $\epsilon>\epsilon_{TE}^c=0.81$ in this case.
The middle and bottom left panels show a 
case for $\epsilon=0.5<\epsilon_{TE}^c$
where a genuine (and stable) TE solution exists, while the corresponding
right panels are for $\epsilon=1>\epsilon_{TE}^c$, past the critical
point where the LP solution degenerates into a TE mode.

\subsection{EP branch}

\begin{figure}[tbp]
\centerline{
\includegraphics[width=8cm,angle=0,clip]{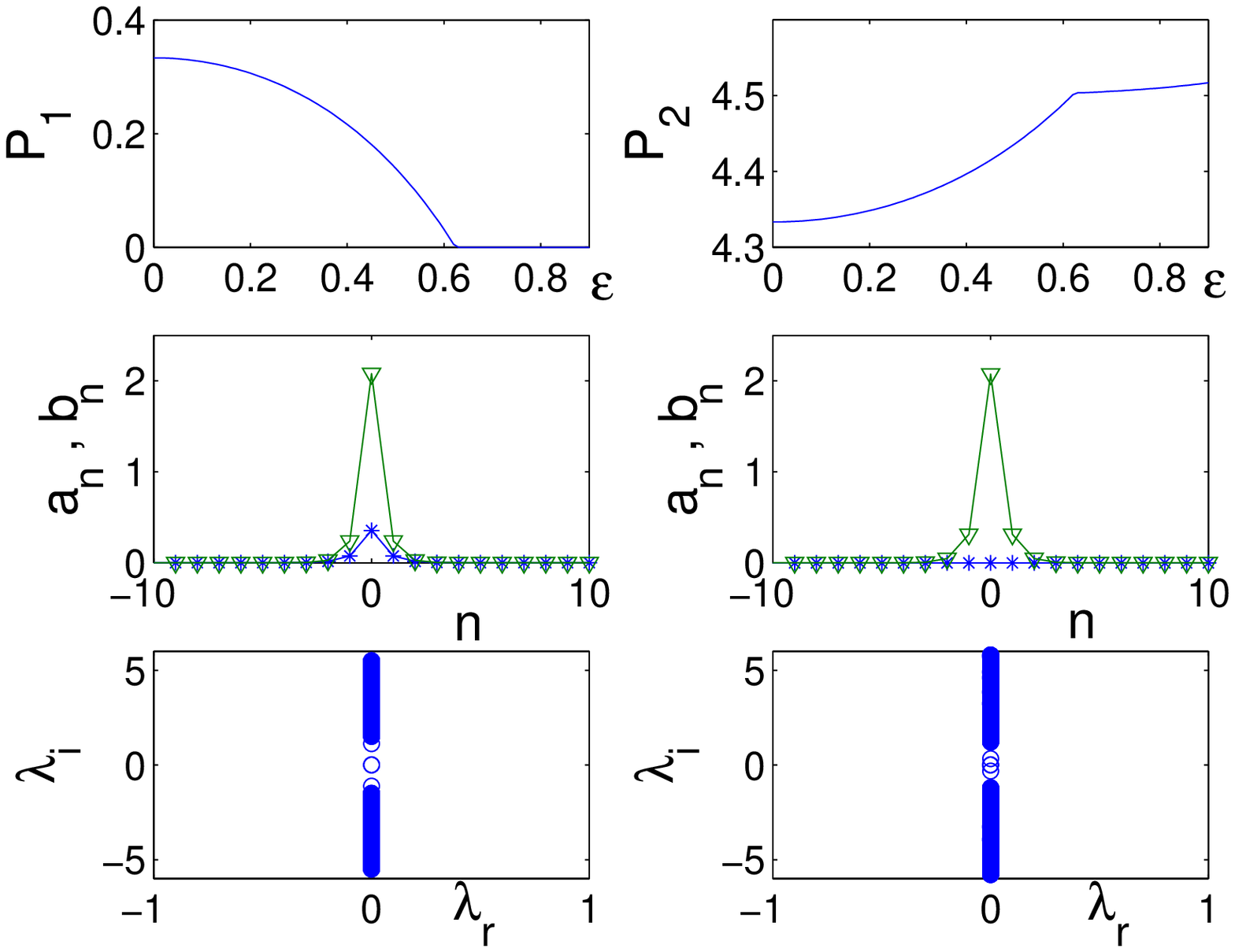}
\includegraphics[width=8cm,angle=0,clip]{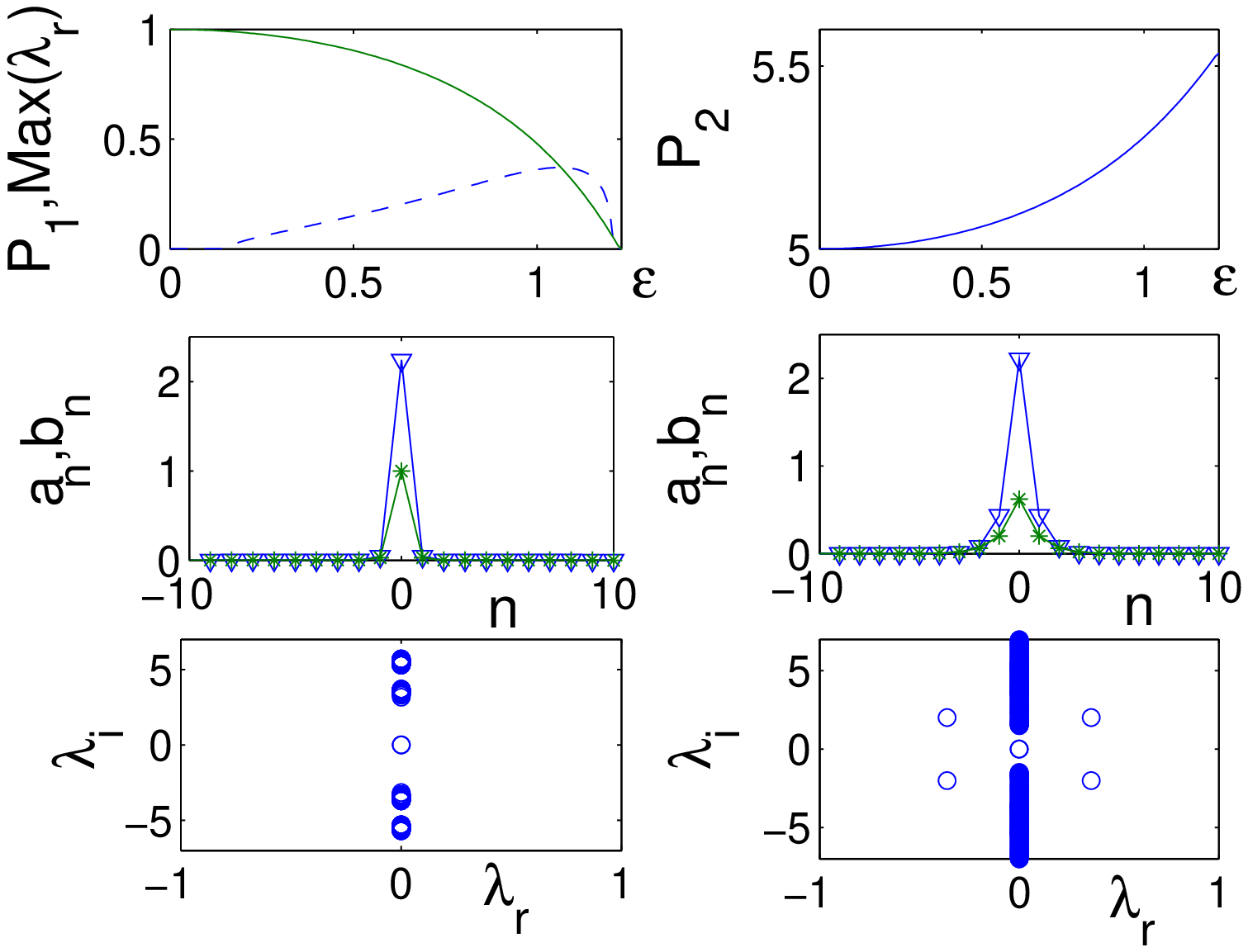}
}
\centerline{
\includegraphics[width=8cm,angle=0,clip]{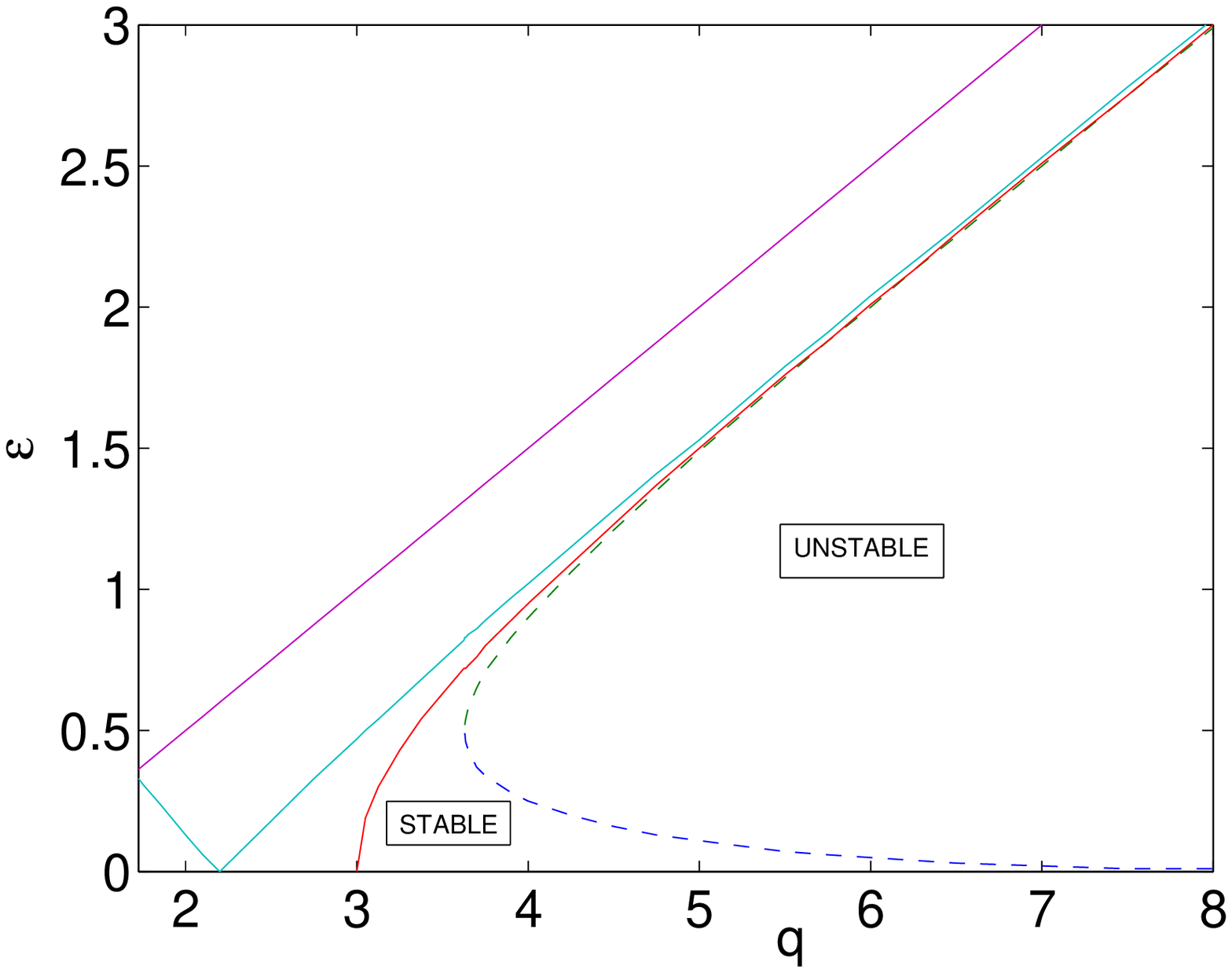}
}
\caption{EP branch: The panels are analogous to
those of Fig. \ref{rfig3} but for the EP branch;
the top right panel, however, shows the non-vanishing 
power of the second component.
Top left panels: $q=3.5$. The middle
and bottom panels correspond to $\epsilon=0.5<\epsilon_{TM}^c$
(left panels) and to $\epsilon=1>\epsilon_{TM}^c$ (right panels).
In the latter case, the solution has already degenerated into
a TM mode. Top right panels: q=4.5. The middle and bottom panels
correspond to the stable case of $\epsilon=0.1$ (left panels)
and the unstable case of $\epsilon=1$ (right panels). The bottom
panel shows the two-parameter diagram for this mode. The region
for $q>3$ and $0<\epsilon<\epsilon_{TM}^c$, where this modes exists
is separated by the dashed line into a stable and an unstable regime.
For comparison the extension of the regions of stability/instability
of the TM mode from Fig. \ref{rfig3} are also included. We note 
that the choice of star symbols (instead of circles as e.g. in Figs.
\ref{rfig2}, \ref{rfig4}) in the spatial profile of $a_n$ is used
to indicate that the imaginary part of $a_n$ is shown.} 
\label{rfig5}
\end{figure}

The case of the EP branch, analyzed in Fig. \ref{rfig5},
is somewhat analogous to that
of the LP branch. For fixed $q$ close to (and larger than) $3$ 
and varying $\epsilon$,
the branch exists and is {\it stable} for $0<\epsilon<\epsilon_{TM}^c$.
For larger values of $\epsilon$, the EP mode degenerates into a
TM mode, as is shown in the top left panel of Fig. \ref{rfig5} for 
$q=3.5$ (cf. with the analogous for the LP case of Fig. \ref{rfig4}). 

However, for $q>3.62$, this phenomenology appears to change and
an expanding (for increasing $q$) interval of oscillatory instability
within the range of existence of the EP branch 
($0<\epsilon<\epsilon_{TM}^c$) seems to arise. To rationalize
this feature, we again turn to the anti-continuum limit. 
Just as in the case of the (TM) branch from which it originates,
the EP branch has a point spectrum eigenfrequency given by Eq. (\ref{req35})
which has a negative Krein signature and upon collision with the continuous
spectrum band of eigenfrequencies will give rise to an instability. Setting
the frequency of Eq. (\ref{req35}) equal to $q-1$, we obtain that
this collision occurs at $q=9$. For lower values of $q$, this ``collision''
will occur for a finite (nonzero) interval of values of $\epsilon$
--see e.g. the top right panels of Fig. \ref{rfig5} for $q=4.5$--, which,
in turn, yields the region of oscillatory instability of the EP mode,
clearly separated from its region of stability by a dashed line 
in the bottom panel of Fig. \ref{rfig5}. 

In conclusion, the EP solutions exist in the regime where the TM
mode is unstable (cf. bottom panels of Figs. \ref{rfig3} and
\ref{rfig5}), illustrated by the curve originating from the point
$(q,\epsilon)=(3,0)$ in the anti-continuum limit. However, the 
EP solution, as is shown in the top panels of the figure,
may (top left) or may not (top right) necessarily be stable
in its region of existence due to the appearance of an 
oscillatory instability for sufficiently large $q$.

\section{Dynamical Evolution of Unstable Branches}

We now examine a series of different instability scenarios
pertaining to the branches of solutions discussed above. In so doing,
we solve Eqs. (\ref{req1})-(\ref{req2}) using a fourth-order 
Runge-Kutta algorithm. To ensure the validity of 
our numerical results, we have monitored the
conservation of
the power of Eq. (\ref{req3a}) and of the Hamiltonian (energy)
of the system: 
\begin{eqnarray}
H= -\sum_n \left(\epsilon (a_n^{\star}a_{n+1}+a_n a_{n+1}^{\star}
+b_n^{\star}b_{n+1}+b_n b_{n+1}^{\star}) + |a_n|^2-|b_n|^2
+ \frac{1}{2} (|a_n|^4+|b_n|^4+A |a_n|^2 |b_n|^2)
+ \frac{B}{2} (a_n^2 (b_n^{\star})^2+b_n^2 (a_n^{\star})^2) \right).
\label{req38}
\end{eqnarray}
The power is conserved to numerical precision, while the Hamiltonian
is conserved to 1 part in $10^8$.  In all cases, $150$ spatial sites 
are used in our computations. 

\begin{figure}[tbp]
\centerline{
\includegraphics[width=8.cm,height=5cm,angle=0,clip]{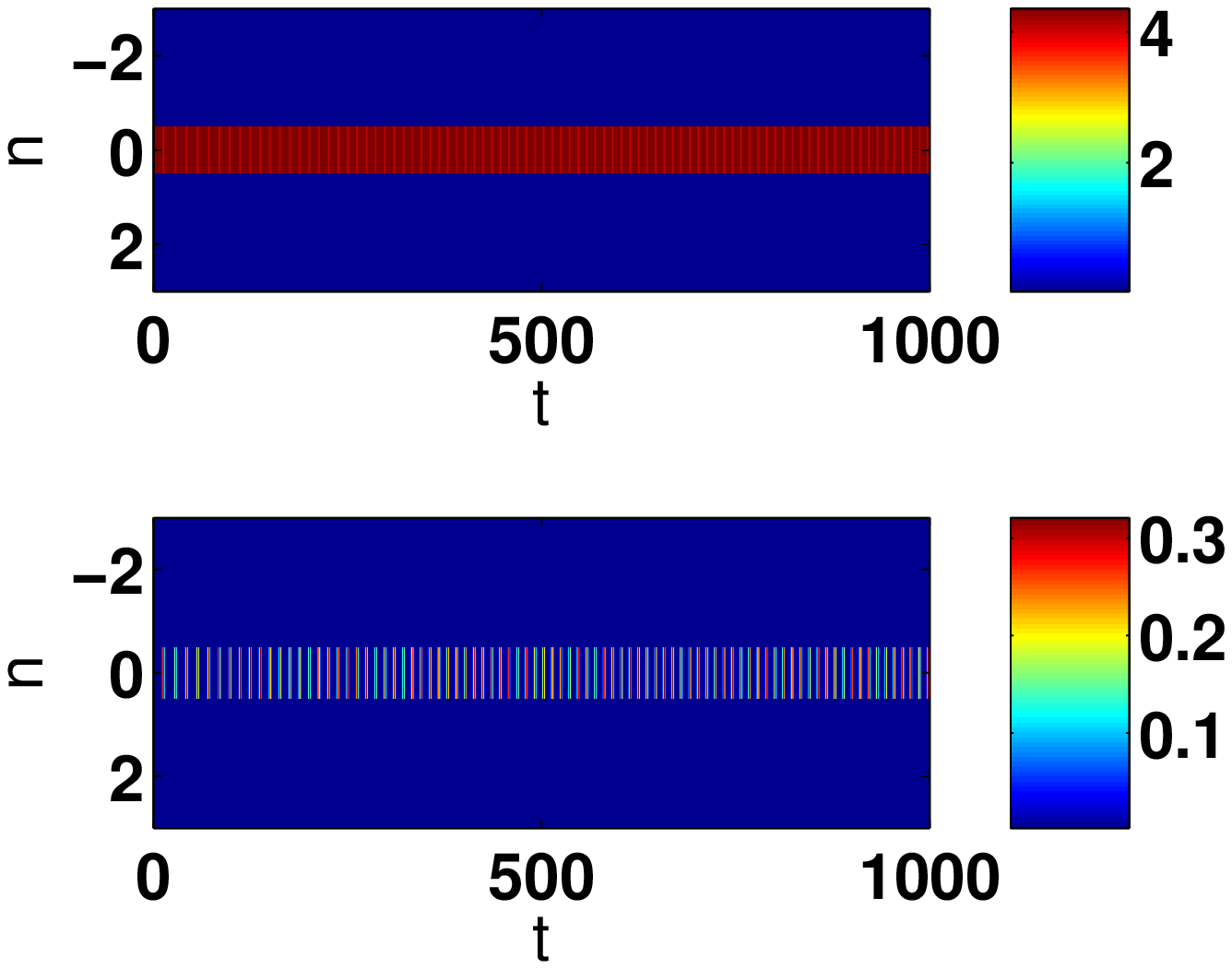}
\includegraphics[width=8.cm,height=5cm,angle=0,clip]{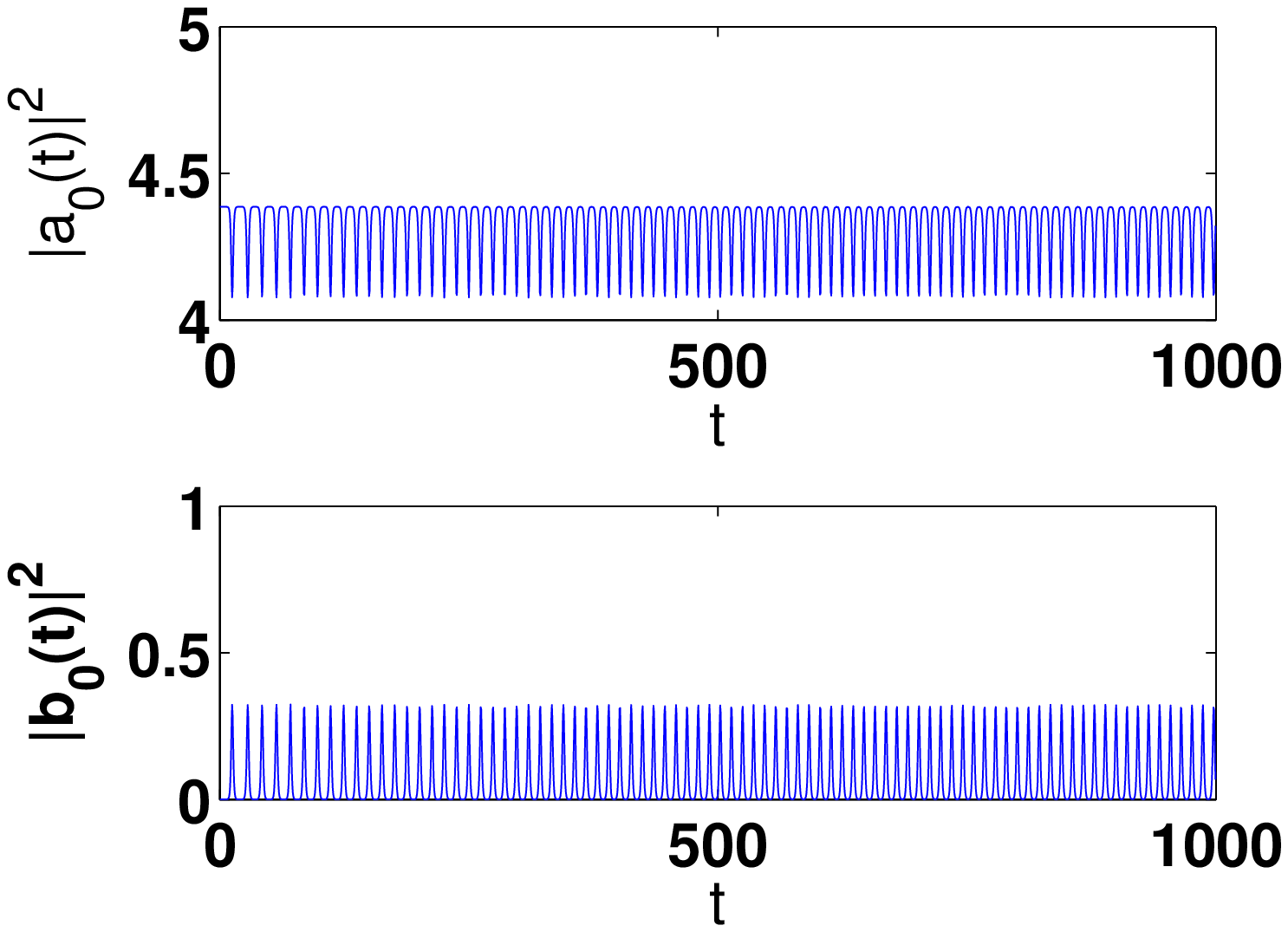}}
\caption{This demonstrates the instability in the TE mode when 
$\epsilon =0.5$ and $q=5.5$. The left panel shows the space-time 
contour plot of the
modulus square of the two fields ($|a_n(t)|^2$ in the top panel and
$|b_n(t)|^2$ in the bottom panel). The right panel shows the evolution
of the (square modulus of the) central site of the configuration as a
function of time.} 
\label{rfig6}
\end{figure}

Our first example, shown in Fig. \ref{rfig6}, 
illustrates the case of $q=5.5$ and $\epsilon=0.5$
for the TE mode. This is a case representative of 
the region of parameter space
(cf. Fig. \ref{rfig2}), where the TE mode is unstable due to a real
eigenvalue pair (while the LP mode is stable). 
The solution actually coincides with the steady TE branch until 
approximately time $10$.
The configuration then deviates from the TE branch and
subsequently develops a breathing oscillation around a stable LP state.

\begin{figure}[tbp]
\centerline{
\includegraphics[width=8.cm,height=5cm,angle=0,clip]{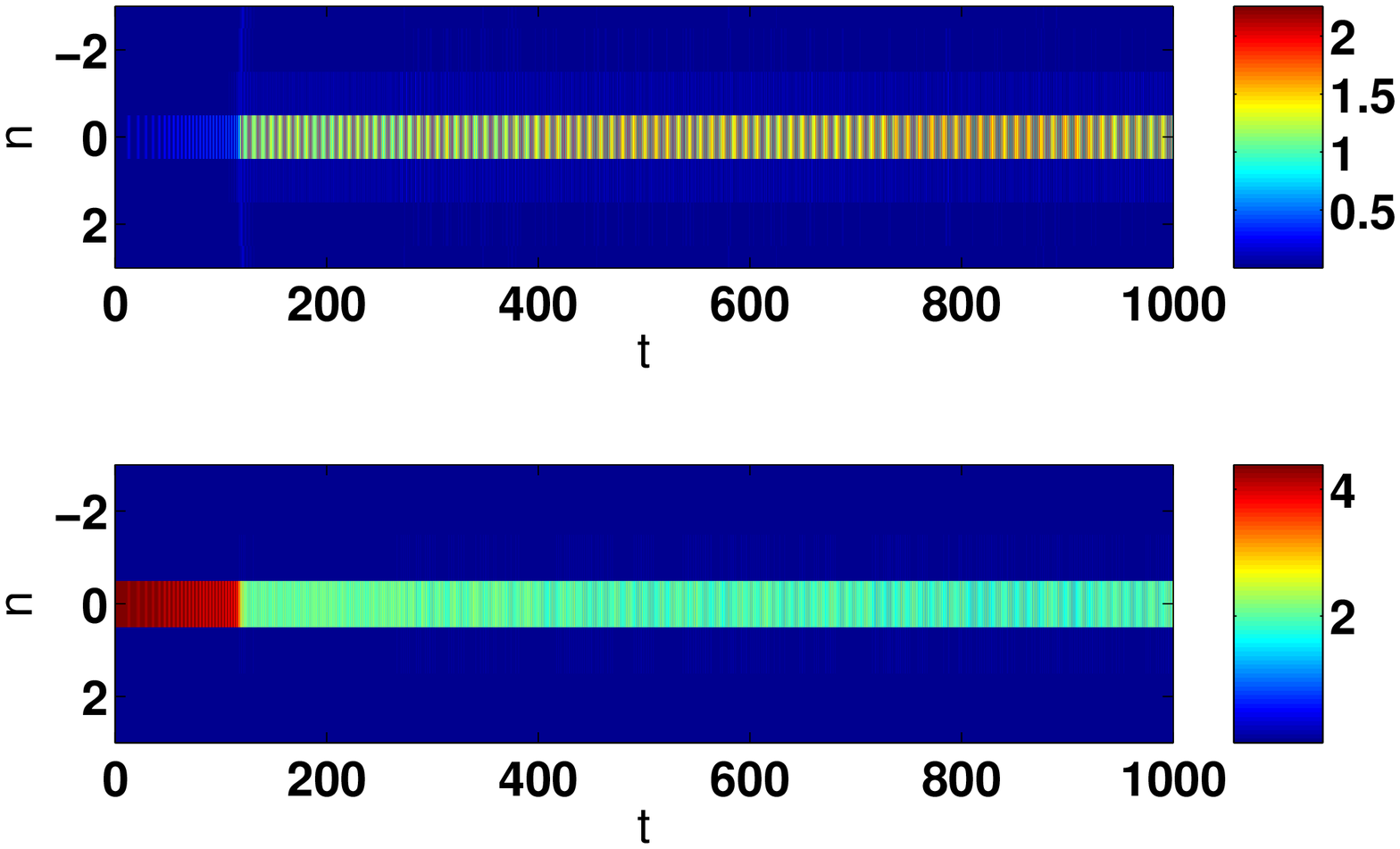}
\includegraphics[width=8.cm,height=5cm,angle=0,clip]{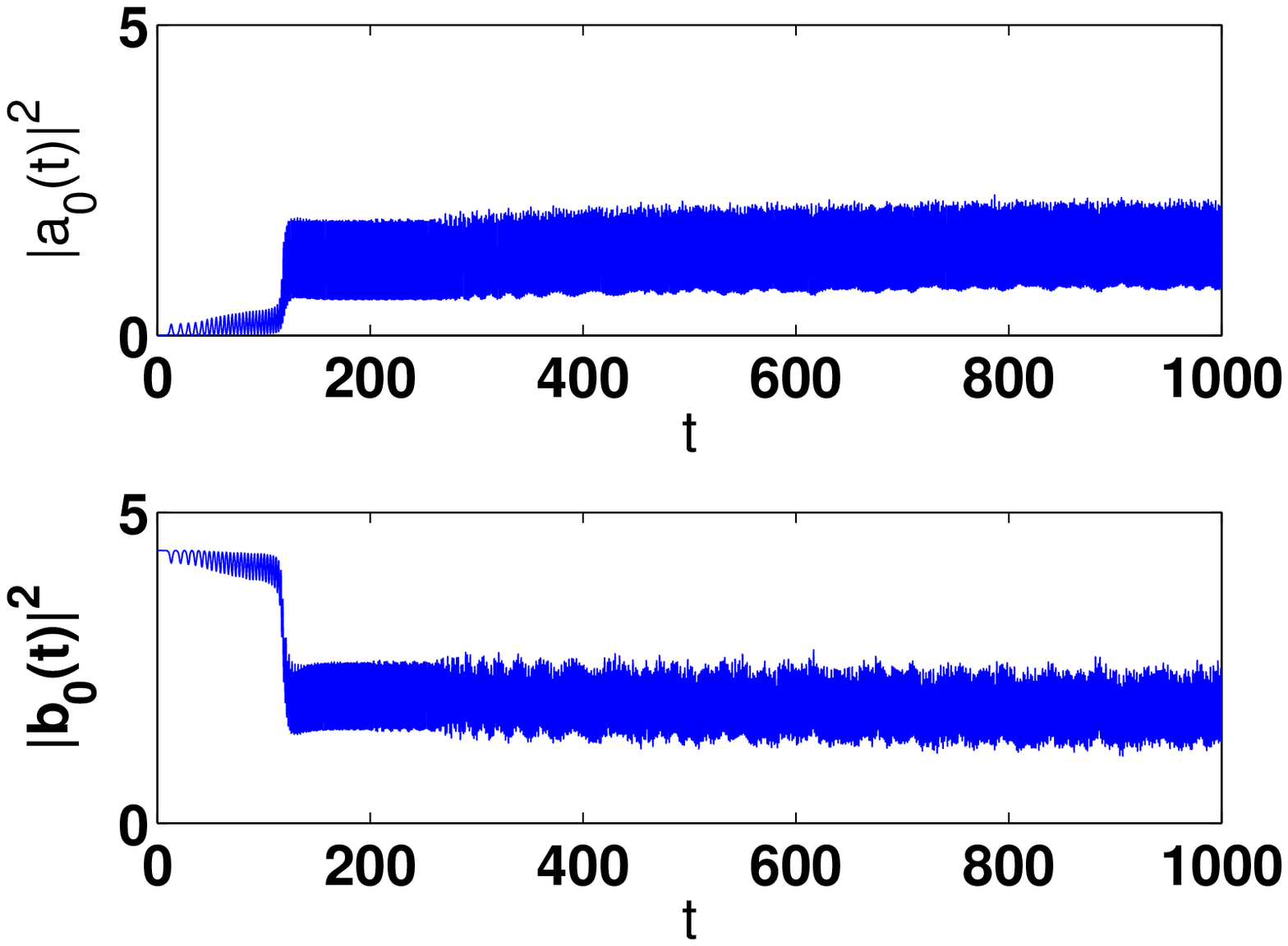}
}
\caption{An instability of the TM mode is demonstrated here with
$\epsilon =0.5$ and $q=3.5$. The left panel again shows the 
space-time contour plot of the 
modulus square of the two fields ($|a_n(t)|^2$ in the top panel and
$|b_n|^2$ in the bottom panel). The right panel shows the evolution
of the (square modulus of the) central site of the configuration as a 
function of time.} 
\label{rfig7}
\end{figure}

The next three cases are concerned with the different scenarios
for the instability of the TM mode. Fig. \ref{rfig7},
with $q=3.5$ and $\epsilon=0.5$, is representative
of the region of parameter space 
where the TM mode is unstable due to a real eigenvalue pair, while the
EP mode is stable. In this case, As in the previous case, 
the solution coincides 
with the steady TM mode up until approximately time $10$. It is then observed  
that for
short times the configuration deviates from the TM branch and oscillates
around the (stable) EP configuration. However, for longer times, 
the system appears to wander away from the latter configuration 
and into a breathing state, where the two components have roughly
equal power and exchange a small fraction of it (almost) periodically.

\begin{figure}[tbp]
\centerline{
\includegraphics[width=8.cm,height=5cm,angle=0,clip]{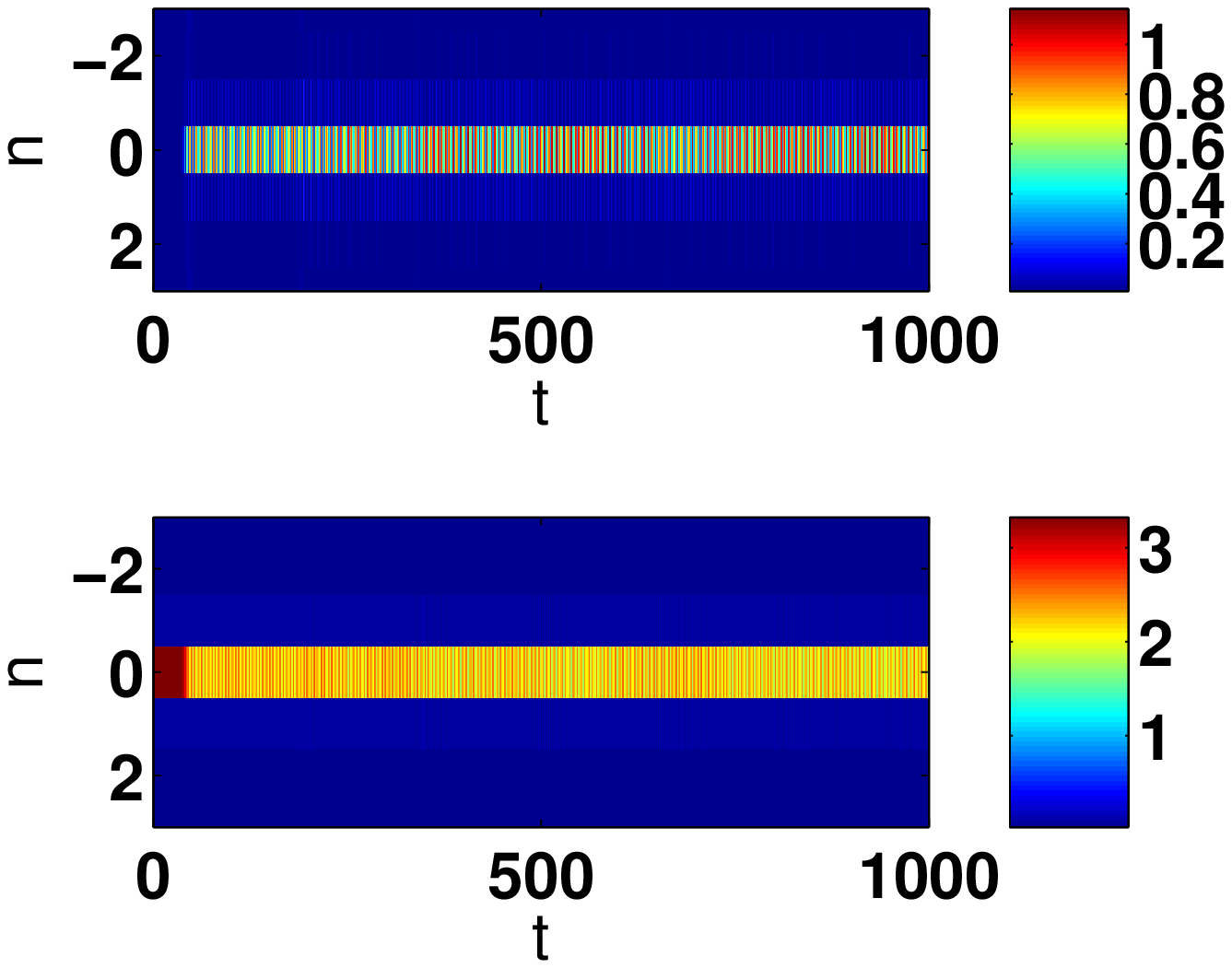}
\includegraphics[width=8.cm,height=5cm,angle=0,clip]{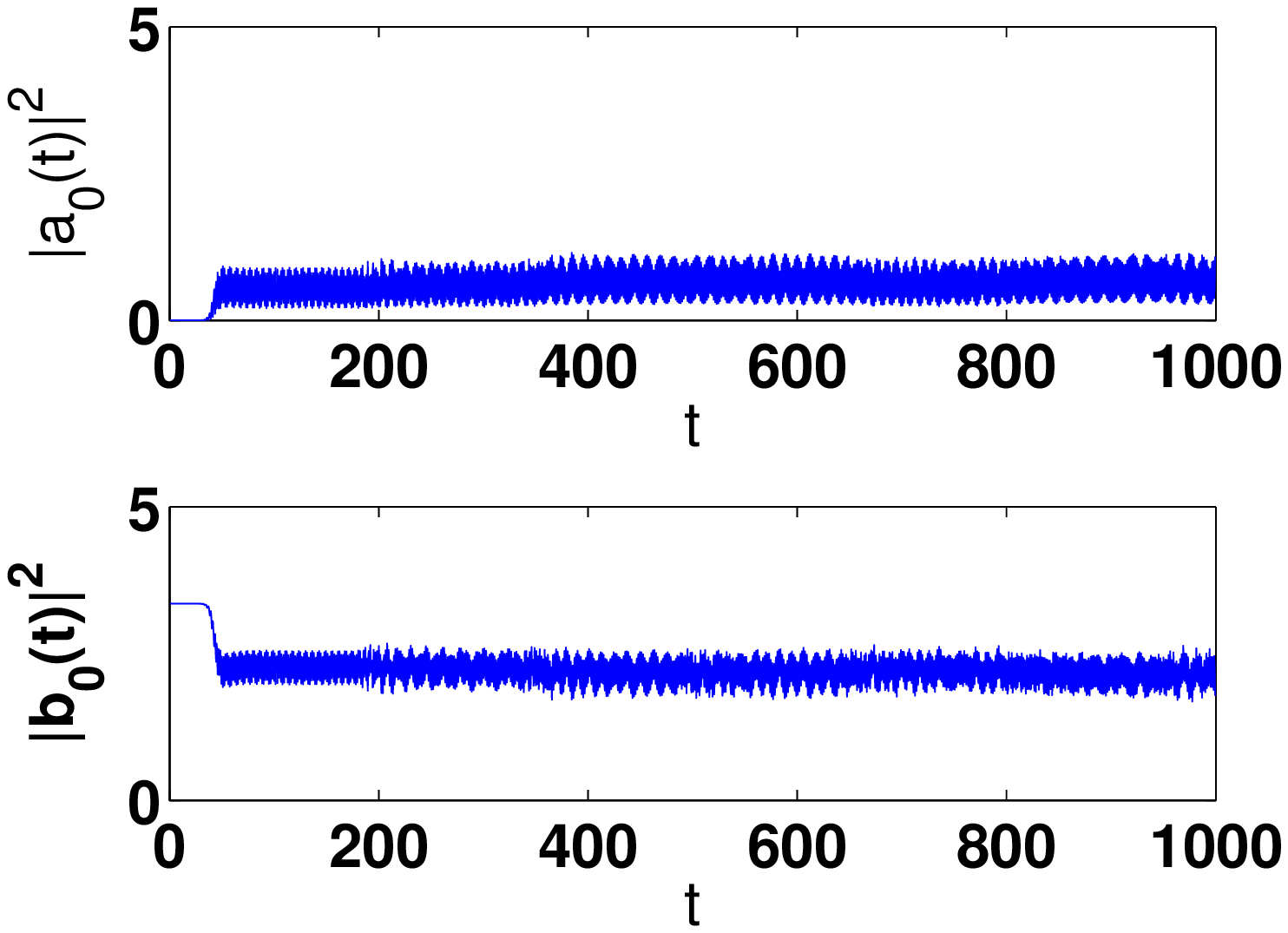}}
\caption{The instability of a TM mode is shown here for
$\epsilon =0.5$ and $q=2.5$, using exactly the same diagnostics
as in Fig. \ref{rfig7}.} 
\label{rfig8}
\end{figure}

The second instability scenario of the TM branch, shown in
Fig. \ref{rfig8},  involves the case
of an eigenvalue quartet as e.g. 
for $q=2.5$ and $\epsilon=0.5$ (cf. Fig. \ref{rfig3}).
In this case, we observe the configuration departing from the 
unstable TM steady state and
resulting into a breathing solution involving both components. In this
and the remaining cases, the departure from the steady state occurs later
than in the first two cases.

Next, we examine the evolution of the instability for the TM
mode in the presence, also, of continuous spectrum instabilities
for $q=2.5$ and $\epsilon=1$ (cf. Fig. \ref{rfig3}). In this case
a different phenomenology occurs. 
The mode is completely destroyed in favor of small amplitude excitations.

\begin{figure}[tbp]
\centerline{
\includegraphics[width=8.cm,height=5cm,angle=0,clip]{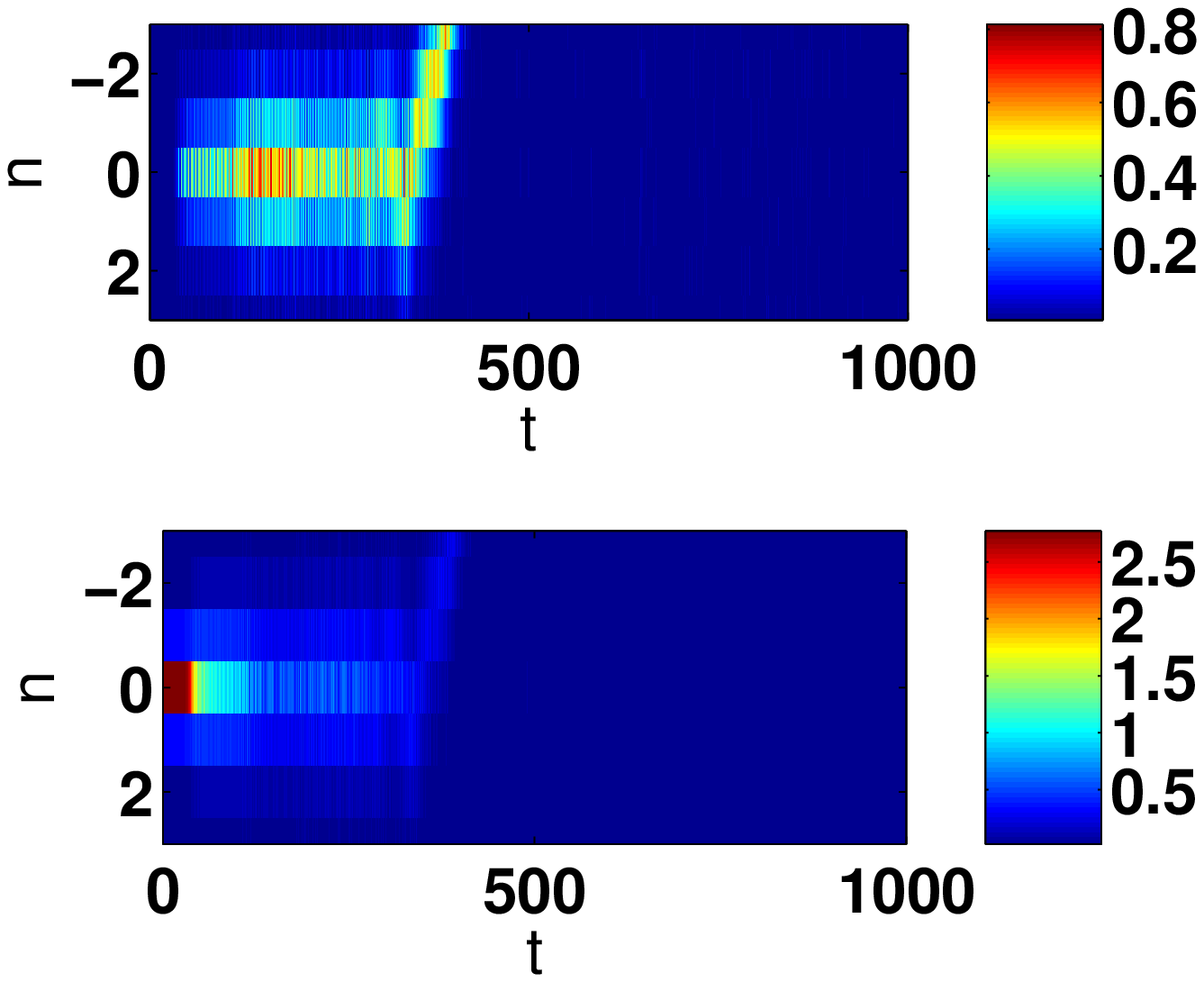}
\includegraphics[width=8.cm,height=5cm,angle=0,clip]{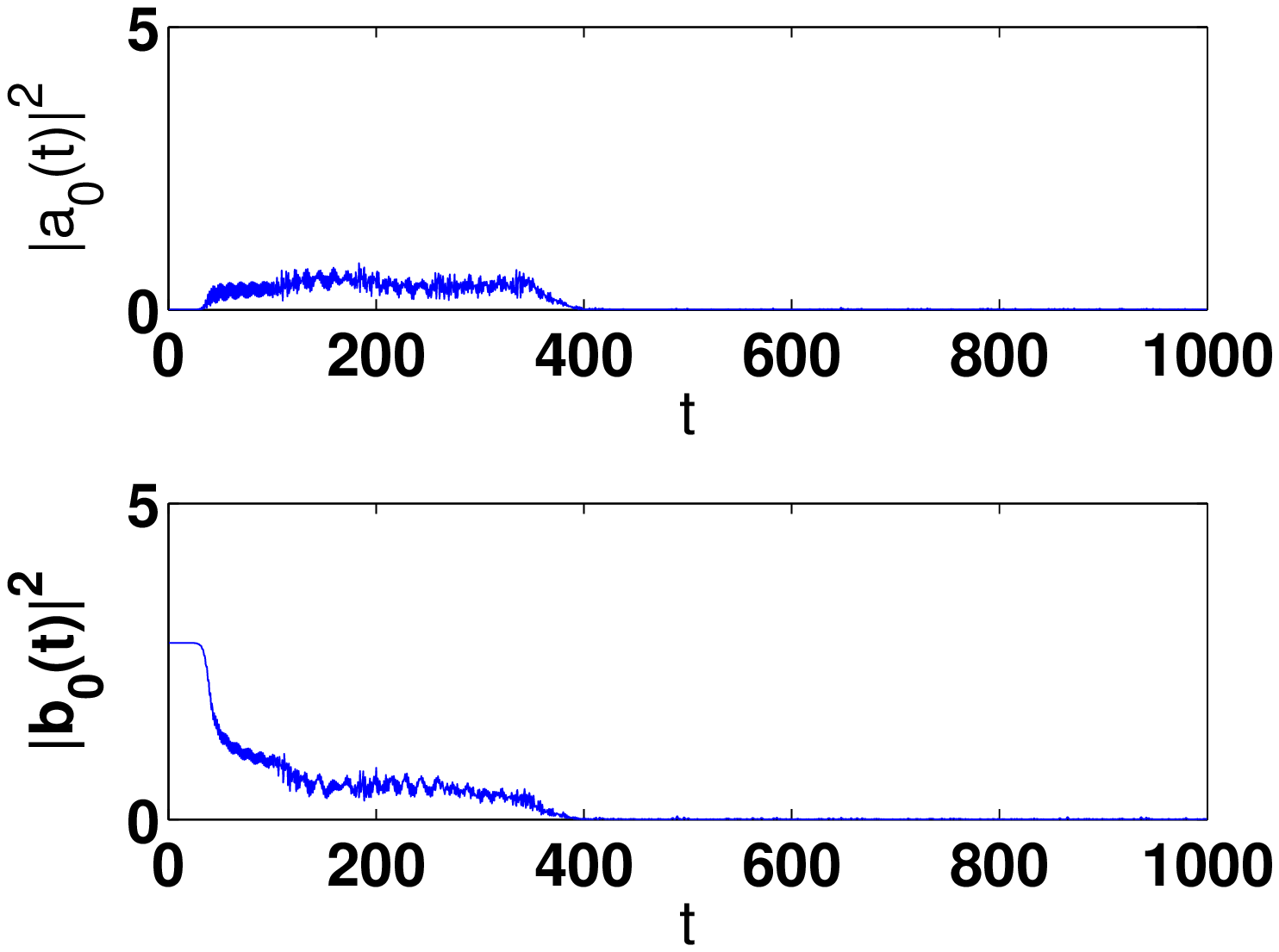}
}
\caption{Same as the previous 2 figures, but for the case of 
$\epsilon =1$ and $q=2.5$. Here the nonlinear mode is completely destroyed
by the instability and results into small amplitude extended waves.} 
\label{rfig9}
\end{figure}


Finally,  we examine the evolution of the instability for an elliptically
polarized mixed mode case within its regime of oscillatory
instability (cf. Fig. \ref{rfig5}), for $\epsilon =0.5$ and $q=5$. 
We observe the configuration departing the steady state and giving
rise to a persistent breathing state involving both components.
                                                                                                
\begin{figure}[tbp]
\centerline{
\includegraphics[width=8.cm,height=5cm,angle=0,clip]{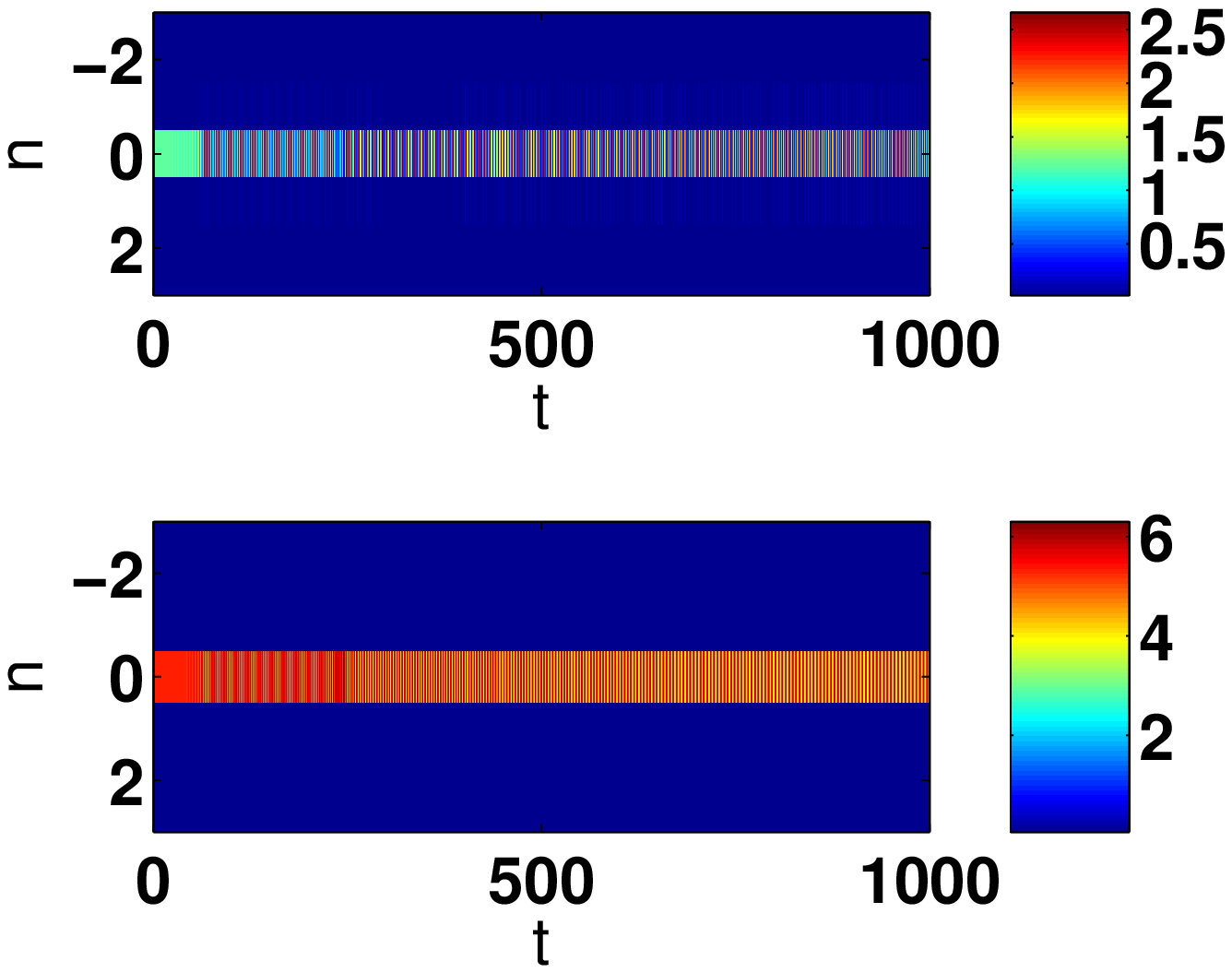}
\includegraphics[width=8.cm,height=5cm,angle=0,clip]{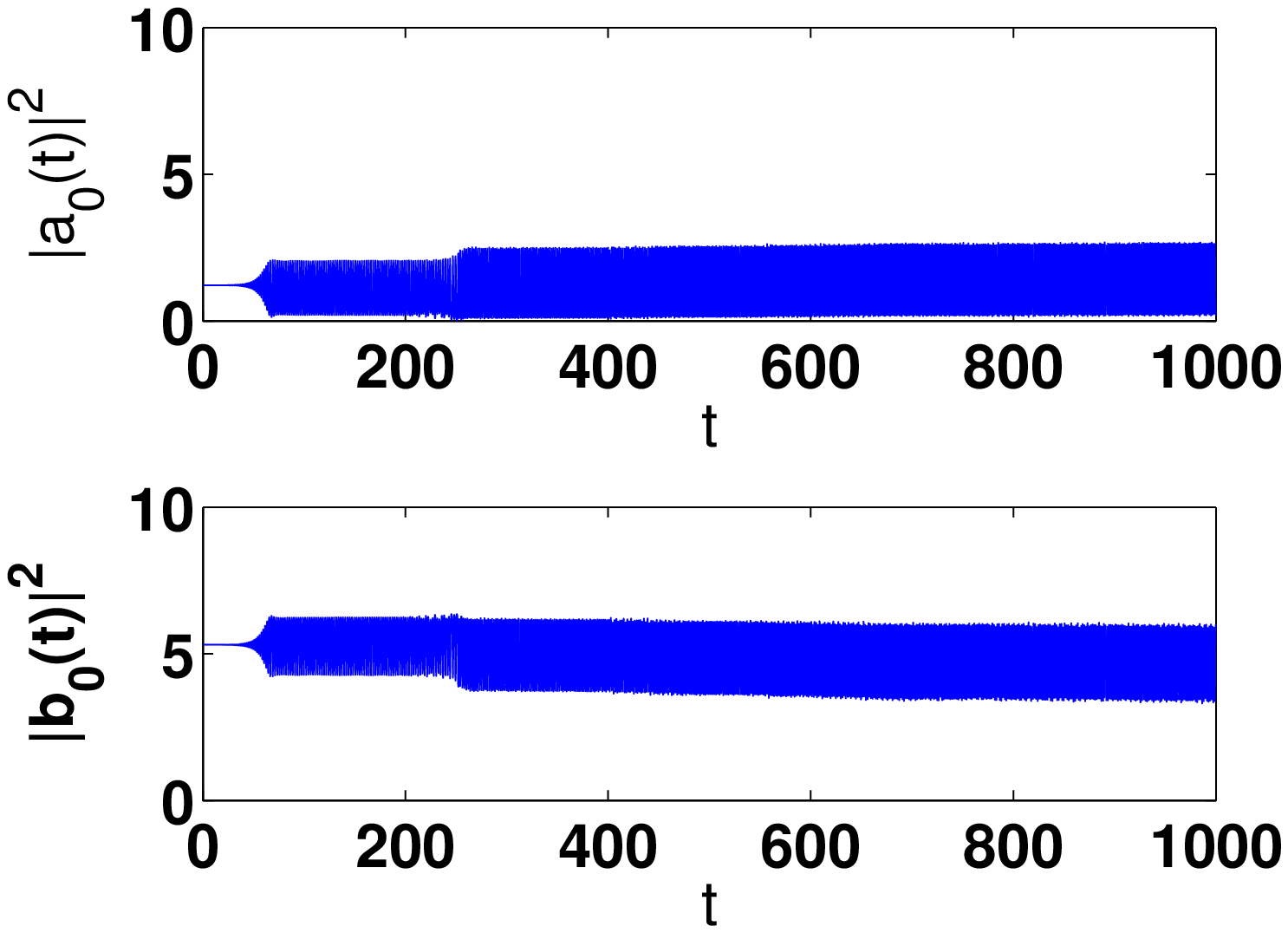}
}
\caption{An unstable elliptically polarized mode with
$\epsilon =0.5$ and $q=5$ is monitored by the same diagnostics
as in the previous figures.}
\label{rfig11}
\end{figure}

\section{Conclusions}

In this paper, we have examined in detail the principal modes arising in the
discrete vector lattices of AlGaAs waveguides, motivated by the recent
experimental investigations of \cite{meier,meier2}. We have constructed
the full two-parameter bifurcation diagrams of all four principal
branches of solutions. These include the transverse electric and
transverse magnetic modes (occupying respectively each of the two
components of the vector lattice), but also the linearly polarized
(two components in or out of phase) and elliptically polarized modes (two 
components out of phase by odd multiples of $\pi/2$). The latter
two modes, emerge from the former two in pairs through a pitchfork
bifurcation that has been completely elucidated through the study
of the system in the anti-continuum limit. In fact, the anti-continuum
limit has provided with a very effective tool to 
{\it analytically} highlight existence
and stability properties of the different branches as well as
potential instabilities. We have continued this 
monoparametric picture (as a function of the propagation constant
of the solutions, but at zero coupling) to finite coupling fleshing
out the regimes of stability and instability, and of existence as
well as of disappearance of the relevant branches of solutions. 
The relevant two-parameter diagrams (Fig. \ref{rfig2} for the TE
mode, Fig. \ref{rfig3} for the TM, Fig. \ref{rfig4} for the LP and
Fig. \ref{rfig5} for the EP mode) constitute the main result
of the present paper. We have, however, complemented our existence
and stability analysis with direct numerical ``experiments'', illustrating
the different instability scenario for the modes of interest and how
they can either lead to the excitation of other branches, or even to 
the complete annihilation of the relevant localized waves.

It is noteworthy that the process of obtaining such a multiparametric
bifurcation picture has allowed us to reveal a considerably deeper
understanding of the system's parameter space, in comparison to 
earlier publications (such as e.g. the theoretical conclusions
of the works of \cite{meier,meier2}). In particular, for instance, 
the limited parameter space examined in the latter setting (where
a monoparametric continuation was used for $\epsilon \simeq 0.921$
and apparently varying $q$ only for $q > 3$) led the authors to infer that the
TM mode is {\it always} unstable and similarly that the bifurcating
EP mode is {\it also} unstable. Here, we have illustrated explicitly
regions of stability of each of these modes in the parameter
space of the system (which definitely exist for lower values of the coupling).

Finally, while the recent investigations of the various principal
branches have offered a rather substantial understanding of their
properties, there are multiple directions of future interest in this
topic that also appear within experimental reach. Such examples consist
e.g. of the study of two-dimensional waveguide arrays and 
of discrete solitons \cite{solitons} and vortices \cite{vortices}
in them. Another such example is the study of more elaborate,
multi-site constructions, such as twisted modes \cite{twisted} even 
in one spatial dimension. While the above directions of examination
of stationary states are of interest in their own right, an additional
dimension of relevance revealed by our direct instability simulations of 
section V pertains to the
systematic study of the existence and stability of breathing 
excitations due to their principal role in the system's dynamics. 
Such investigations are currently in progress
and will be reported  in future publications.

\end{document}